\documentclass[journal ]{new-aiaa}
\usepackage[utf8]{inputenc}
\usepackage{textcomp}

\usepackage{graphicx}
\usepackage{amsmath}
\usepackage[version=4]{mhchem}
\usepackage{siunitx}
\usepackage{longtable,tabularx}
\usepackage{float}

\title{ {\small \it under consideration for publication in AIAA Journal} \\ \vspace{4mm}
A Lagrange multiplier-based optimal control technique for streak attenuation in high-speed boundary layers}

\author{Omar Es-Sahli \footnote{Lecturer, Department of Mechanical Engineering; AIAA Member.} and Adrian Sescu\footnote{Associate Professor, Department of Aerospace Engineering; AIAA Associate Fellow.}}

\affil{Mississippi State University, Mississippi State, MS 39762}

\author{M. Z. A. Koshuriyan \footnote{Lecturer, Department of Mathematics; AIAA Member.}}

\affil{University of York, Department of Mathematics, Heslington, York YO10 5DD, UK}

\author{Yuji Hattori\footnote{Professor, Institute of Fluid Science.} and Makoto Hirota\footnote{Assistant Professor, Institute of Fluid Science.\\
This article presents the final results of a previously published conference paper presented at the AIAA Aviation 2021 Virtual Forum \url{https://doi.org/10.2514/6.2021-2945}.}}

\affil{Tohoku University, 2 Chome-1-1 Katahira, Aoba Ward, Sendai, 980-8577, Japan}

\begin{document}

\maketitle

\begin{abstract}


High-amplitude free stream turbulence and surface roughness elements can excite a laminar boundary layer flow sufficiently to cause streamwise oriented vortices to develop. These vortices resemble elongated streaks having alternate spanwise variations of the streamwise velocity. Downstream, the vortices `wobble' through an inviscid secondary instability mechanism and, ultimately, transition to turbulence. We formulate an optimal control algorithm to suppress the growth rate of the streamwise vortex system. Considering a high Reynolds number asymptotic framework, we reduce the full compressible Navier-Stokes equations to the nonlinear compressible boundary region equations (NCBRE). We then implement the method of Lagrange multipliers via an appropriate transformation of the original constrained optimization problem into an unconstrained form to obtain the disturbance equations in the form of the adjoint compressible boundary region equations (ACBRE) and corresponding optimality conditions. Numerical solutions of the ACBRE approach for high-supersonic and hypersonic flows reveal a significant reduction in the kinetic energy and wall shear stress for all considered configurations. We present contour plots to demonstrate the qualitative effect of increased control iterations. Our results indicate that the primary vortex instabilities gradually flatten in the spanwise direction thanks to the ACBRE algorithm.

\end{abstract}

\section{Introduction}
Flow control techniques are an integral part of fluid mechanics research as they offer possible optimization and improvement in the performance and operation of mechanical systems involving fluids. 
Automobiles, aircraft, and marine vehicles all offer  {possible avenues for} implementation of such control methods to determine the optimal configuration of wings, surface/edge geometry, engine inlet, nozzle design, and many other components to improve  {design metrics}.
The (mathematically) optimized design could potentially increase the lift to drag ratio on an aerodynamic surface, delay/accelerate the transition to turbulence, postpone flow separation, enhance mixing or alter the morphology of the vortices embedded within the turbulence field. 

The main goal behind research into controlling transitional or fully-developed turbulent boundary layers is to reduce the energy carried by aligned vortical structures in the streamwise direction. 
These structures appear in the form of alternate high- and low-velocity streaks originating in the near-wall region (e.g., Kendall \cite{Kendall}, Matsubara \& Alfredsson \cite{Matsubara}, or Landahl \cite{Landahl}). For boundary layers over flat plates or wings, such streaky flows occur when the upstream roughness height exceeds a critical value (e.g., Choudhari \& Fischer \cite{choudhari}, White \cite{white1}, White \textit{et al.} \cite{white2}, Goldstein \textit{et al.} \cite{Goldstein1,Goldstein2,Goldstein3}, and Wu \& Choudhari \cite{wu1}) or when the amplitude of the free-stream disturbances surpasses a certain threshold. (e.g., Kendall \cite{Kendall}, Westin \textit{et al.}  \cite{Westin}, Matsubara \& Alfredsson \cite{Matsubara}, Leib \textit{et al.} \cite{Leib}, Zaki \& Durbin \cite{Zaki}, Goldstein \& Sescu \cite{Goldstein4}, and Ricco \textit{et al.} \cite{Ricco}).

G\"{o}rtler vortices refer to streamwise elongated streaks in a boundary layer flow over a concave surface. They develop due to the imbalance between the centrifugal forces and the wall-imposed pressure gradient (e.g., G\"{o}rtler \cite{Gortler}, Hall \cite{hall1,hall2,hall3}, Swearingen \& Blackwelder \cite{Swearingen}, Malik \& Hussaini \cite{Malik}, Saric \cite{Saric}, Li \& Malik \cite{Li}, Boiko \textit{et al.} \cite{Boiko}, Wu \textit{et al.} \cite{wu}, or Sescu \textit{et al.} \cite{Sescu2,Sescu3}, Es-Sahli \textit{et al.} \cite{Es-Sahli1}, Ren \& Fu \cite{Ren}, Dempsey \textit{et al.} \cite{Dempsey}, Xu \textit{et al.} \cite{Xu}). 
The higher the wall curvature, the more rapid the vortex formation is, and therefore, the quicker transition occurs. Low to medium wall curvature also induce vortex formation, and consequently altering the primary flow and causing the laminar flow to break down.

Xu \textit{et al.} \cite{Xu_2} investigated the influence of the curvature, turbulence level, and pressure gradient on the development of streaks and G\"{o}rtler vortices in a boundary layer over a flat or concave wall. They found that an adverse/favorable pressure gradient causes the G\"{o}rtler vortices to saturate earlier/later, but at a lower/higher amplitude than that in the case of a zero-pressure-gradient. On the other hand, for the same pressure gradient and at low levels of the free stream vortical disturbances (FSVD), the vortices saturate earlier and at a higher amplitude as G\"{o}rtler number increases.
Raising FSVD intensity reduces the effects of the pressure gradient and curvature. At a high FSVD level of 14 \%, the curvature has no impact on the vortices, while the pressure gradient only influences the saturation intensity.
	
Chen \textit{et al.} \cite{Chen} used linear stability theory (LST) and direct numerical simulation (DNS) to investigate the stability and transition of G\"{o}rtler vortices in a hypersonic boundary layer at M = 6.5. 
They used blowing and suction to excite G\"{o}rtler vortices with different spanwise wavelengths L3, L6, and L9 (3, 6, and 9 mm, respectively). Their results showed, among other findings, that for L3, the sinuous-mode instability dominates the breakdown process, whereas, for L6, the varicose-mode instability is most dangerous. For L9, sinuous- and varicose-mode instabilities appear to be of equal importance, with the former controlling the near-wall region and the latter appearing on the top of the mushroom structures. We will show subsequently that our optimal control method is very effective in controlling the near-wall sinuous-mode instability, making it very suitable for small and large spanwise wavelength cases where this instability is dominant. 

Like many other types of instabilities, streamwise-oriented vortices (and their accompanying streaks) occur in various engineering applications such as the flow over turbo-machinery blades{,} and flow in the proximity to the walls of wind tunnels or turbofan engine intakes.
Not only do these instabilities contribute to transition, but they also induce a significant increase in noise in supersonic and hypersonic wind tunnels, which can cause interference with the measurements in the test section (Schneider \cite{Schneider}). 
Thus comparing and validating wind tunnel measurements to real-flight conditions becomes challenging.
The objective function for a control algorithm in this context would be to reduce the streamwise vortex energy, which would delay nonlinear breakdown and the transition  {onset}. 
However, since the transient part of the primary instability is responsible for the growth of three-dimensional disturbances (and  their subsequent breakdown), the applied control strategy must restrict transient modes growth.

Wall transpiration control techniques of boundary layers can be applied via localized suction and blowing regions underneath low-and high-velocity streaks. 
The net result is a decrease in the spanwise variation of the streamwise velocity and, therefore, a corresponding reduction in the number and strength of the bursting events. 

A more efficient way of achieving these results is by applying active wall control methods. 
In the recent literature, such methods have been utilized in turbulent channel flow as a technique to reduce skin friction drag (see Choi \textit{et al.} \cite{Choi}). 
Choi \textit{et al.} \cite{Choi} conducted direct numerical simulations of an active-control approach based on wall transpiration velocity by placing sensors in a sectional plane parallel to the wall. 
This technique achieved approximately a 25\% reduction in frictional drag. But from a practical point of view, it is challenging, if not impossible, to install sensors in the flow as they  {could potentially} amplify existing disturbances or create new ones. Thus, to avoid interference issues, Choi \textit{et al.} \cite{Choi} investigated the same control algorithm with sensors placed at the wall that inform (i.e., feed into) the leading term in the Taylor series expansion of the vertical velocity component near the wall. However, this approach resulted only in a 6\% reduction. Koumoutsakos \cite{Koumoutsakos1, Koumoutsakos2} implemented a similar feedback control algorithm informed by flow quantities at the wall. Taking the vorticity flux components as inputs to the control algorithm resulted in a significant drop in skin friction (approximately 40\%). 

Lee \textit{et al.} \cite{Lee1} derived new suboptimal feedback control laws based on blowing and suction to manipulate the flow structures in the proximity to the wall using surface pressure or shear stress distribution (the reduction in the frictional drag was in the range of 16-20\%). 
Observing that the opposition control technique is more effective at low Reynolds number turbulent wall flows, Pamies \textit{et al.} \cite{Pamies} proposed the utilization of blowing only at high Reynolds numbers, and by doing so, they obtained a significant reduction in the skin-friction drag for these flows. Recently, Stroch \textit{et al.} \cite{Stroh} conducted a comparison between the opposition control applied in the framework of turbulent channel flow and a spatially developing turbulent boundary layer. They found that the rates of frictional drag reduction are approximately similar in both cases. The paper by Kim \cite{Kim1} reviews the technical issues and limitations associated with the opposition control type and we refer the interested reader to it.

In a linear, full-state optimal control theory framework, Hogberg \textit{et al.} \cite{Hogberg1} reported the first successful re-laminarization of a $Re_{\tau}=100$ turbulent channel flow by applying zero mass flux blowing and suction at the wall. They showed that the information available in the linearized equations is sufficient to construct linear controllers able to re-laminarize a turbulent wall flow, but this may be limited to low Reynolds number flows. 

In addition to numerical methods studies, several experimental studies used blowing and suction to control the disturbances in laminar or turbulent boundary layer flows. We briefly mention  {a few of these} here. 
Gad-el-Hak \& Blackwelder \cite{Gad1} used continuous or intermittent suction to eliminate artificially generated instabilities in a flat-plate boundary layer flow. Experiments of Myose \& Blackwelder \cite{Myose} used the same idea to delay the breakdown of G\"{o}rtler vortices.
Jacobson \& Reynolds \cite{Jacobson} developed a new type of actuation based on a vortex generator to control disturbances induced by a cylinder along the wall normal-direction and unsteady boundary layer streaks generated using pulsed suction. Regarding the latter, the actuation significantly reduced the spanwise gradients of the streamwise velocity, which are one of the main driving forces of secondary instabilities (see Swearingen \& Blackwelder \cite{Swearingen}). Experiments of Lundell \& Alfredsson \cite{Lundell} controlled streamwise velocity streaks in a channel flow by using localized suction regions downstream of the inflow. 
This approach delays secondary instabilities and ultimately the onset of transition.

Many studies employed optimal control for laminar or turbulent boundary layer flows. 
There have been several studies that considered the implementation of optimal control for shear flows; see for example, Gunzburger \cite{Gunzburger} or the more recent review by Luchini \& Bottaro \cite{Luchini}
(note that the latter is in a slightly different context).
Other studies aimed at controlling disturbances developing in laminar or turbulent boundary layers {have been numerous}; ({\it e.g.} Bewley \& Moin \cite{Bewley}, Joslin \textit{et al.} \cite{Joslin}, Cathalifaud \& Luchini\cite{Cathalifaud}, Corbet \& Bottaro \cite{Corbett}, Hogberg \textit{et al.} \cite{Hogberg1}, Zuccher \textit{et al.} \cite{Zuccher}, Cherubini \textit{et al} \cite{Cherubini}, Lu \textit{et al.} \cite{Lu}, Sescu \& Afsar \cite{Sescu6}, Sescu \textit{et al.} \cite{Sescu7}). 

Bewley \& Moin \cite{Bewley} applied an optimal control method based on blowing and suction to turbulent channel flows. They reported a 17\% frictional drag reduction as a result of this scheme. Cathalifaud \& Luchini \cite{Cathalifaud} used the same optimal control approach to reduce the energy of disturbances in a boundary layer flow over a flat plate and a concave surface. The study of Zuccher \textit{et al.} \cite{Zuccher} discussed and tested an optimal and robust control strategy in the framework of steady three-dimensional disturbances in the form of streaks that formed in a boundary layer flow over a flat plate. They established the optimal control method on an adjoint-based optimization technique to first find the optimal state for a given set of initial conditions and then determine the worst initial conditions for optimal control. 
Lu \textit{et al.} \cite{Lu} derived an optimal control algorithm within the linearized unsteady boundary region equations.
As we shall see in §.$2$, these equations are the asymptotic reduced form of the Navier-Stokes equations under assumptions of low frequency and low streamwise wavelength. Their study aimed at controlling both streaks developing in flat-plate boundary layer flows and G\"{o}rtler vortices evolving along concave surfaces. Cherubini \textit{et al.} \cite{Cherubini} applied a nonlinear optimal control strategy with blowing and suction, starting with the full Navier-Stokes equations and using the method of Lagrange multipliers to determine the maximum decrease of the disturbance energy.

Papadakis, Lu \& Ricco \cite{Papadakis} derived a closed-loop optimal control technique based on wall transpiration in the framework of a flat plate laminar boundary layer excited by freestream disturbances. They split the optimal control into two sequences{: of which are} obtained by marching the corresponding equations in the forward and backward directions respectively. 
They found that the feedback sequence is more effective than the feed-forward sequence. Xiao \& Papadakis \cite{Xiao} derived an optimal control algorithm in the framework of the full Navier-Stokes equations and Lagrange multipliers. The study aimed at delaying transition in a flat plate boundary layer excited by freestream vortical disturbances based on blowing and suction.

Although researchers have made significant progress in the last decades toward understanding the phenomenology behind the formation and influence of streamwise vortices and streaks in incompressible boundary layers, research into the compressible regime still remains modest.
Moreover, bypass-transition at high upstream flow speeds remains largely unexplored. In this paper, we consider the control of streamwise vortices in boundary layers by conducting a parametric study of various flow parameters within a self-consistent asymptotic mathematical framework.
The paper is an extension of the incompressible flow analysis of Sescu \& Afsar \cite{Sescu6} into the compressible regime. They implemented an optimal control approach to limit the growth of G\"{o}rtler vortices developing in an incompressible laminar boundary layer flow over a concave wall. But the adjoint boundary region equations and the associate optimality condition were based on the incompressible boundary region equations.
They showed that the optimal control algorithm (employed wall deformation and velocity transpiration as flow control parameters) is very effective in reducing the amplitude of the G\"{o}rtler vortices, especially for the control based on wall displacement. In this paper, on the other hand, we derive the ACBRE for compressible flows together with the optimality condition for the NCBRE.
The latter are the high Reynolds number extension of the compressible Navier-Stokes equations. We introduce appropriate control variables (i.e., wall transpiration velocity) with wall shear stress as the cost functional.

\section{Scaling and governing equations}

\subsection{Scalings}

All dimensional spatial coordinates $(x^*,y^*,z^*)$ are normalized by the spanwise wavelength of the freestream disturbance $\lambda^*$ set to 0.5cm, while the dependent variables by their respective freestream values. The pressure {field is} normalize{d} by the dynamic pressure. {Hence:}
	\begin{align}
	    &\bar{t} = \frac{t^*}{\lambda^*/U_{\infty}^*}; \hspace{4mm} \bar{x} = \frac{x^*}{\lambda^*}; \hspace{4mm}
	    \bar{y} = \frac{y^*}{\lambda^*}; \hspace{4mm} \bar{z} = \frac{z^*}{\lambda^*} \nonumber\\
	    &\quad \bar{u} = \frac{u^*}{U_{\infty}^*}; \hspace{4mm} \bar{v} = \frac{v^*}{U_{\infty}^*}; \hspace{4mm}
	    \bar{w} = \frac{w^*}{U_{\infty}^*}; \hspace{4mm} \bar{\rho} = \frac{\rho^*}{\rho_{\infty}^*} \nonumber \\
	    &\quad \bar{p} =  \frac{p^*-p_{\infty}^*}{\rho_{\infty}^*V_{\infty}^{*2}}; \hspace{4mm} \bar{T} = \frac{T^*}{T_{\infty}^*};
	    \hspace{4mm} \bar{\mu} = \frac{\mu^*}{\mu_{\infty}^*}; \hspace{4mm} \bar{k} = \frac{k^*}{k_{\infty}^*} \nonumber
	\end{align}
where $u^*$, $v^*$, and $w^*$ are the velocity components in the streamwise, wall-normal, and spanwise directions, respectively. $\rho^*$, $p^*$,  $T^*$, $\mu^*$, and $k^*$ represent the dimensional density,  pressure,  temperature,  dynamic viscosity, and thermal conductivity, respectively. 
	
	We define the Reynolds number based on the spanwise wavelength, the Mach number, and the Prandtl number as
	\begin{eqnarray}
	R_{\lambda} = \frac{\rho_{\infty}^* U_{\infty}^* \lambda^*}{\mu_{\infty}^*}, \hspace{5mm}
	M_\infty = \frac{U_{\infty}^*}{c_{\infty}^*}, \hspace{5mm}
	Pr = \frac{\mu_{\infty}^* C_p}{k_{\infty}^*} \nonumber
	\end{eqnarray}
    where $C_p$ is the specific heat at constant pressure and $\mu_{\infty}^*$, $c_{\infty}^*$ and $k_{\infty}^*$ are the freestream dynamic viscosity, speed of sound, and thermal conductivity, respectively. 
    
    For boundary layer flows over curved surfaces, we define the {$\mathcal{O}(1)$} global G\"{o}rtler number as
	\begin{eqnarray}
	G_{\lambda} = \frac{R_{\lambda}^2 \lambda^*}{r^*} \nonumber
	\end{eqnarray}
	where $r^*$ is the radius of the curvature (set to 1.5m).
	
	\subsection{Compressible Navier-Stokes equations (N-S)}
	
	 For a full compressible flow, the primitive form of the Navier-Stokes equations in non-dimensional variables are as follows

	
	\begin{eqnarray}
	\label{Eq_1}
	\frac{D \bar{\rho}}{D t} 
	+ \rho \left( \frac{\partial \bar{u}}{\partial \bar{x}} + \frac{\partial \bar{v}}{\partial \bar{y}} + \frac{\partial \bar{w}}{\partial \bar{z}} \right) = 0
	\end{eqnarray}
	
	
	\begin{equation}
		\label{Eq_2}
	\bar{\rho} \frac{D \bar{u}}{D \bar{t}}
	= -\frac{\partial \bar{p}}{\partial \bar{x}} 
	+ \frac{1}{R_\lambda}\frac{\partial}{\partial \bar{x}} \left[ \frac{2}{3} \mu \left( 2\frac{\partial \bar{u}}{\partial \bar{x}} - \frac{\partial \bar{v}}{\partial \bar{y}} - \frac{\partial \bar{w}}{\partial \bar{z}} \right) \right]
	+ \frac{\partial}{\partial \bar{y}} \left[ \mu \left( \frac{\partial \bar{u}}{\partial \bar{y}} + \frac{\partial \bar{v}}{\partial \bar{x}} \right) \right]
	+ \frac{\partial}{\partial \bar{z}} \left[ \mu \left( \frac{\partial \bar{w}}{\partial \bar{x}} + \frac{\partial \bar{u}}{\partial \bar{z}} \right) \right]
	\end{equation}
	
	
	\begin{equation}
		\label{Eq_3}
	\bar{\rho} \frac{D \bar{v}}{D \bar{t}}
	= -\frac{\partial \bar{p}}{\partial \bar{y}} 
	+ \frac{1}{R_\lambda}\frac{\partial}{\partial \bar{y}} \left[ \frac{2}{3} \mu \left( 2\frac{\partial \bar{v}}{\partial \bar{y}} - \frac{\partial \bar{u}}{\partial \bar{x}} - \frac{\partial \bar{w}}{\partial \bar{z}} \right) \right]
	+ \frac{\partial}{\partial \bar{x}} \left[ \mu \left( \frac{\partial \bar{v}}{\partial \bar{x}} + \frac{\partial \bar{u}}{\partial \bar{y}} \right) \right]
	+ \frac{\partial}{\partial \bar{z}} \left[ \mu \left( \frac{\partial \bar{v}}{\partial \bar{z}} + \frac{\partial \bar{w}}{\partial \bar{y}} \right) \right]
	\end{equation}
	
	
	\begin{equation}
		\label{Eq_4}
	\bar{\rho} \frac{D \bar{w}}{D \bar{t}}
	= -\frac{\partial \bar{p}}{\partial \bar{z}} 
	+ \frac{1}{R_\lambda}\frac{\partial}{\partial \bar{z}} \left[ \frac{2}{3} \mu \left( 2\frac{\partial \bar{w}}{\partial \bar{z}} - \frac{\partial \bar{u}}{\partial \bar{x}} - \frac{\partial \bar{v}}{\partial \bar{y}} \right) \right]
	+ \frac{\partial}{\partial \bar{x}} \left[ \mu \left( \frac{\partial \bar{w}}{\partial \bar{x}} + \frac{\partial \bar{u}}{\partial \bar{z}} \right) \right]
	+ \frac{\partial}{\partial \bar{y}} \left[ \mu \left( \frac{\partial \bar{v}}{\partial \bar{z}} + \frac{\partial \bar{w}}{\partial \bar{y}} \right) \right]
	\end{equation}
	
	
	\begin{align}
		\label{Eq5}
	& \bar{\rho} \frac{D \bar{T}}{D \bar{t}}
	= 
	\frac{1}{Pr R_\lambda} \left[ \frac{\partial}{\partial \bar{x}} \left( k \frac{\partial \bar{T}}{\partial \bar{x}} \right) + \frac{\partial}{\partial \bar{y}} \left( k \frac{\partial \bar{T}}{\partial \bar{y}} \right) + \frac{\partial}{\partial \bar{z}} \left( k \frac{\partial \bar{T}}{\partial \bar{z}} \right) \right]  \nonumber  \\
	&\quad - (\gamma - 1) M_{\infty}^2 \left[ p \left( \frac{\partial \bar{u}}{\partial \bar{x}} + \frac{\partial \bar{v}}{\partial \bar{y}} + \frac{\partial \bar{w}}{\partial \bar{z}} \right)
	-\frac{2}{3} \mu \left( \frac{\partial \bar{u}}{\partial \bar{x}} + \frac{\partial \bar{v}}{\partial \bar{y}} + \frac{\partial \bar{w}}{\partial \bar{z}} \right)^2 \right] \\
	&\quad + (\gamma - 1) M_{\infty}^2 \frac{\mu}{R_\lambda} \left[ 2\left( \frac{\partial \bar{u}}{\partial \bar{x}} \right)^2 + 2\left( \frac{\partial \bar{v}}{\partial \bar{y}} \right)^2 + 2\left( \frac{\partial \bar{w}}{\partial \bar{z}} \right)^2
	+ \left( \frac{\partial \bar{u}}{\partial \bar{y}} + \frac{\partial \bar{v}}{\partial \bar{x}} \right)^2
	+ \left( \frac{\partial \bar{w}}{\partial \bar{x}} + \frac{\partial \bar{u}}{\partial \bar{z}} \right)^2
	+ \left( \frac{\partial \bar{v}}{\partial \bar{z}} + \frac{\partial \bar{w}}{\partial \bar{y}} \right)^2 \right]   \nonumber
	\end{align}
	where 
	\begin{equation}
		\label{Eq_6}
	\frac{D}{D\bar{t}} = \frac{\partial}{\partial \bar{t}} + \bar{u} \frac{\partial}{\partial \bar{x}} + \bar{v} \frac{\partial}{\partial \bar{y}} + \bar{w} \frac{\partial}{\partial \bar{z}}
	\end{equation}
	is the substantial derivative (for what follows, we consider the steady-state case for the N-S equations, i.e. $\partial / \partial \bar{t} = 0$). The pressure $p$, the temperature $T$  and the density $\rho$ of the fluid are combined in the equation of state in non-dimensional form, $\bar{p} = \bar{\rho} \bar{T} / \gamma M_{\infty}^2$, under the assumption of non-chemically-reacting flows. Other notations include the dynamic viscosity $\mu$, the thermal conductivity $k$, and the free-stream Mach number $M_{\infty}=U_{\infty}^*/c_{\infty}^*$, which is otherwise $\mathcal{O}(1)$. We write $\mu$ and $k$ as a function of the temperature using the power law of viscosity in the form
	
	\begin{equation}
	\mu = T^b;  \hspace{6mm}
	k = \frac{C_p \mu}{Pr}
	\end{equation}
	where $b=0.76$ (Ricco \& Wu \cite{Ricco1}), $C_1 = 1.458 \times 10^{-6}$, $C_2 = 110.4$, $C_p = \gamma R / (\gamma - 1)$, $\gamma = 1.4$, and $Pr = 0.72$ for air.
	
	\subsection{Nonlinear Compressible Boundary-Region Equations (NCBRE)}
	
\begin{figure}[hbt!]
 \begin{center}
 \includegraphics[width=\linewidth]{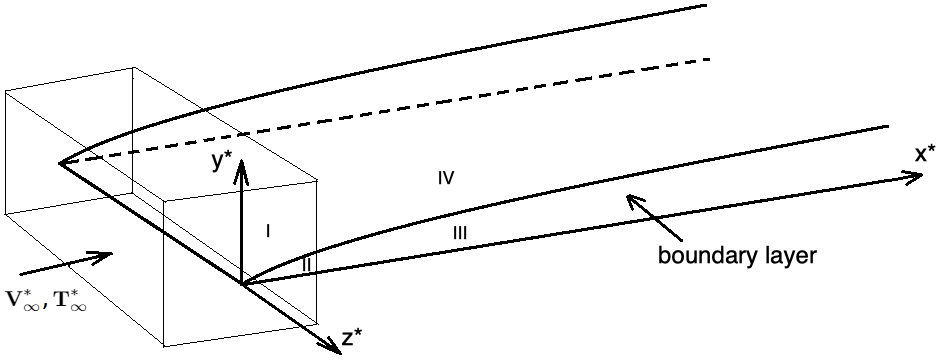}
 \end{center}
 \caption{The flow domains illustrating the asymptotic structure and the four regions of the flow domain. The box distinguishes regions I and II with I inside the box and outside the boundary layer and II inside the box and inside the boundary layer}
 \label{f1}
\end{figure}

We consider a compressible flow of uniform velocity $U_{\infty}^*$ and temperature $T_{\infty}^*$ past a flat or curved surface. The air is treated as a perfect gas so that the sound speed in the free-stream $c_{\infty}^* =\sqrt{\gamma R T_{\infty}^*}$, where $\gamma$ = 1.4 is the ratio of the specific heats, and $R = 287.05 N m/(kg K)$ is the universal gas constant; Mach number is assumed to be of order one. The asymptotic structure of the flow is composed of four regions as in Leib \textit{et al.} \cite{Leib}, Ricco \& Wu \cite{Ricco} or Marensi \textit{et al.} \cite{Marensi} (see figure \ref{f1}). 

Close to the leading edge{,} and outside the boundary layer{,} is region I, where the flow is inviscid and the disturbances are small perturbations of the base flow. Region II is the boundary layer close to the leading edge with a thickness much smaller than the freestream disturbances spanwise wavelength, $\lambda$.
{The} linearized boundary region equations {now} govern the perturbation {field solutions in which} the spanwise {diffusion} is of the same order of magnitude as that in the wall-normal direction. Region III is the {fully} viscous region governed by the nonlinear compressible boundary region equations (NCBRE) where the boundary layer thickness is of the same order of magnitude as $\lambda$. Region IV is inviscid since the viscous effects are negligible. In the {latter} region, the displacement effect due to the increased thickness of the viscous layer influences the flow at leading order. 

We derive the NCBRE from the full steady-state compressible N-S equations. Thanks to the characteristics of the flow in region III, we re-scale the streamwise distance and time co-ordinate at which the vortex system forms by the following $\mathcal{O}(1)$ variables: $x = \bar{x}/R_{\lambda}$, and the time as $t = \bar{t}/R_{\lambda}$. Note that the distance in the wall-normal and spanwise directions are the same, $y = \bar{y}$, $z = \bar{z}$. Another thing to mention is that, in this region, the crossflow velocity component is small compared to the streamwise component, and pressure variations are negligible. Appropriate dominant balance considerations suggest that the dependent variables in this region must also re-scale as follows:
	\begin{align}
	&u = \bar{u}; \hspace{4mm}  v = \bar{v} /R_{\lambda}; \hspace{4mm}  w = \bar{w} /R_{\lambda}; \hspace{4mm} 
	\rho = \bar{\rho}; \hspace{4mm} \nonumber \\ 
	&\quad p = \bar{p} /R_{\lambda}^2; \hspace{4mm} 
	T = \bar{T}; \hspace{4mm}
	\mu = \bar{\mu}; \hspace{4mm}  k = \bar{k} \nonumber
	\end{align}
Inserting the above re-scaled dependent and independent variables into the full unsteady continuity (\ref{Eq_1}), Navier Stokes  (\ref{Eq_2})-(\ref{Eq_4}) and energy equation (\ref{Eq5}) gives the following leading order equations in curvilinear coordinates (that we refer to as the NCBRE):

	
	\begin{equation}\label{neq1}
		\textbf{V} \cdot \nabla \rho 
		+ \rho \nabla \cdot \textbf{V} = 0
	\end{equation}
	
	
	\begin{equation}\label{neq2}
		\rho \textbf{V} \cdot \nabla u
		= 
		\nabla_{c} \cdot \left( \mu \nabla_{c} u \right)
	\end{equation}
	
	
	\begin{equation}\label{neq3}
		\rho \textbf{V} \cdot \nabla v + G_\lambda u^2
		= 
		-\frac{\partial p}{\partial y} 
		+ \frac{\partial}{\partial y} \left[ \frac{2}{3} \mu \left( 3\frac{\partial v}{\partial y} - \nabla \cdot \textbf{V} \right) \right] 
		+ \frac{\partial}{\partial x} \left( \mu \frac{\partial u}{\partial y} \right) 
		+ \frac{\partial}{\partial z} \left[ \mu \left( \frac{\partial v}{\partial z} + \frac{\partial w}{\partial y} \right) \right]
	\end{equation}
	
	
	\begin{equation}\label{neq4}
		\rho \textbf{V} \cdot \nabla w
		= 
		-\frac{\partial p}{\partial z} 
		+ \frac{\partial}{\partial z} \left[ \frac{2}{3} \mu \left( 3\frac{\partial w}{\partial z} - \nabla \cdot \textbf{V} \right) \right]
		+ \frac{\partial}{\partial x} \left( \mu \frac{\partial u}{\partial z} \right)
		+ \frac{\partial}{\partial y} \left[ \mu \left( \frac{\partial v}{\partial z} + \frac{\partial w}{\partial y} \right) \right]
	\end{equation}
	
	
	\begin{equation}\label{neq5}
		\rho \textbf{V} \cdot \nabla T
		= 
		\frac{1}{Pr} \nabla_{c} \cdot \left( k \nabla_{c} T \right)
		+(\gamma - 1) M_{\infty}^2 \mu \left[ \left( \frac{\partial u}{\partial y} \right)^2
		+ \left( \frac{\partial u}{\partial z} \right)^2 \right]
	\end{equation}
	where $\textbf{V}$ is the velocity vector and $\nabla_{c}$ is the crossflow nabla operator:
	\begin{equation}
	\textbf{V} = u \vec{i} + v \vec{j} + w \vec{k}; \hspace{6mm} \nabla_{c} = \frac{\partial}{\partial y} \vec{j} + \frac{\partial}{\partial z} \vec{k},
	\end{equation}
	and the temporal and streamwise derivatives are absent because they remain $\mathcal{O}(1/R_\lambda)$ and are therefore asymptotically small for the vortex flow in region III. 
	The effect of the wall curvature is contained in the term involving the global G\"{o}rtler number, $G_\lambda = \mathcal{O}(1)$.
	
	This set of equations is parabolic in the streamwise direction and elliptic in the spanwise direction. Appropriate initial/upstream and boundary conditions are necessary to close the problem; these conditions will be the same as those used by Ricco \& Wu \cite{Ricco}. We solve the NCBRE using the numerical algorithm developed in Es-Sahli \textit{et al.} \cite{Es-Sahli2}.
	
	\section{Optimal control problem in the nonlinear regime}\label{s3}
	
	While it is common for an optimal flow control problem to be formulated in the framework of a dynamical system (usually described by a set of equations that are parabolic in time), {in the present case} we replace the time direction by the $x$-direction owing to the parabolic character of the NCBRE in the stream-wise direction, which necessitates stream-wise marching to obtain a solution. 
	
	\subsection{Generic optimal control formalism}
	To fix ideas, we first write equations (\ref{neq1})-(\ref{neq5}) in the generic and more compact form
	
	\begin{equation}\label{oo}
	\mathcal{G}(\textbf{q}) = 0,
	\end{equation}
	for brevity, with initial and boundary conditions
	
	\begin{equation}
	\textbf{q}(0,y,z) = \textbf{q}_0(y,z)
	\end{equation}
	\begin{equation}
	\textbf{q}(x,0,z) = \phi, \hspace{3mm}  \lim_{y \rightarrow \infty} \textbf{q}(x,y,z) = \textbf{q}_B(x,z),
	\end{equation}
	along the wall-normal direction, and periodic or symmetry boundary conditions in the span-wise direction, $z$. In equation (\ref{oo}),
	$\mathcal{G}()$ is the NCBRE differential operator in abstract notation, $\textbf{q} = (\rho,u,v,w,T)$ is the vector of state variables, $\phi$ is the control variable associated with the boundary conditions ({\it e.g.}, the transpiration velocity at the wall, $v_w$), $\textbf{q}_0(y,z)$ represents the upstream or initial condition in $x=0$, and $\textbf{q}_B$ is a given function that specifies the boundary condition at infinity. We define an objective (or cost) functional as
	
	\begin{equation}\label{zz}
	\mathcal{J}(\textbf{q},\phi) = \mathcal{E}(\textbf{q}) 
	+ \sigma \left(\| \phi_x \|^{\beta_2} + \| \phi \|^{\beta_2} \right),
	\end{equation}
	where $\mathcal{E}(\textbf{q})$ is a specified target function to be minimized ({\it e.g.}, the energy of the disturbance, or the wall shear stress; the latter is considered in this study), the second term on the right hand side of (\ref{zz}) is a penalization term depending on the norm of the control variable (usually, this quantity place a constraint on the magnitude of the admissible control variable, since it cannot increase or decrease indefinitely), $\sigma$ and $\beta$ are given constants, and subscript $x$ denotes derivative with respect to $x$. The norm $\| \|$ in equation (\ref{zz}) is {now} associated with an appropriate inner product of two complex functions, $f$ and $g$, defined as 
	
	\begin{equation}\label{aaa}
	\langle f, g \rangle = \int_{0}^{X_t} f^* g dX
	\end{equation}
	in the space $[0,X_t]$, with $X_t$ being the terminal streamwise location (the star in (\ref{aaa}) denotes complex conjugate).
	
	A common approach to transform a (nonlinear) constrained optimization problem into an unconstrained problem is by using the method of Lagrange multipliers (see, for example, Joslin \textit{et al.} \cite{Joslin}, Gunzburger \cite{Gunzburger}, Zuccher \textit{et al.} \cite{Zuccher}). To this end, we consider the Lagrangian
	
	\begin{equation}
	\mathcal{L}(\textbf{q},\phi,\textbf{q}^a) = \mathcal{J}(\textbf{q},\phi) - \langle \mathcal{G}(\textbf{q}), \textbf{q}^a \rangle,
	\end{equation}
	where $\textbf{q}^a$ is the vector of Lagrange multipliers $(\rho^a,u^a,v^a,w^a,T^a)$, also known as the adjoint vector. In other words, the Lagrange multipliers are introduced in order to transform the minimization of $\mathcal{J}(\textbf{q},\phi)$ under the constraint $\mathcal{G}(\textbf{q}) = 0$ into the unconstrained minimization of $\mathcal{L}(\textbf{q},\phi,\textbf{q}^a)$. The unconstrained optimization problem can be formulated as: 
	
	{\it Find the control variable $\phi$, the state variables $\textbf{q}$, and the adjoint variables $\textbf{q}^a$ such that the Lagrangian $\mathcal{L}(\textbf{q},\phi,\textbf{q}^a)$ is a stationary function, that is}
	
	\begin{equation}\label{ab}
	\delta \mathcal{L} = \frac{\partial \mathcal{L}}{\partial \textbf{q}} \delta \textbf{q}
	+ \frac{\partial \mathcal{L}}{\partial \phi} \delta \phi
	+ \frac{\partial \mathcal{L}}{\partial \textbf{q}^a} \delta \textbf{q}^a = 0
	\end{equation}
	where 
	
	\begin{equation}
	\frac{\partial \mathcal{L}}{\partial a} \delta a = \frac{\mathcal{L}(a+\epsilon \delta a) - \mathcal{L}(a)}{\epsilon}
	\end{equation}
	represents directional differentiation in the generic direction $\delta a$. All directional derivatives in (\ref{ab}) must vanish, providing different sets of equations: 
	
	\begin{itemize}
		
		\item adjoint BRE equations are obtained by taking the derivative with respect to $\textbf{q}$,
		
		\begin{eqnarray}\label{ii1}
		\frac{\partial \mathcal{L}}{\partial \textbf{q}} = 0  \hspace{4mm} \Rightarrow   \hspace{4mm}   \mathcal{G}^a(\textbf{q}^a) = 0  
		\end{eqnarray}
		
		\item optimality conditions are obtained by taking the derivatives with respect to $\phi$,
		
		\begin{eqnarray}\label{ii2}
		\frac{\partial \mathcal{L}}{\partial \phi} = 0  \hspace{4mm} \Rightarrow   \hspace{4mm}   \mathcal{O}(\textbf{q}^a,\textbf{q},\phi) = 0 
		\end{eqnarray}
		
		\item the original BRE equations are obtained by taking the derivative with respect to $\textbf{q}^a$,
		
		\begin{eqnarray}\label{ii3}
		\frac{\partial \mathcal{L}}{\partial \textbf{q}^a} = 0  \hspace{4mm} \Rightarrow   \hspace{4mm}   \mathcal{G}(\textbf{q},\psi) = 0.
		\end{eqnarray}
		
	\end{itemize}
	
	Equations (\ref{ii1})-(\ref{ii3}) form the optimal control system that can be utilized to determine the optimal states and the control variables. One can note that the stationarity of the Lagrangian with respect to the adjoint variables $\textbf{q}^a = (\rho^a,u^a,v^a,w^a,T^a)$ essentially yields the original state equations, while the stationarity with respect to the state variables $\textbf{q} = (\rho,u,v,w,T)$ yields the adjoint equations that depend on the state variables. The relationship between the state variables and the adjoint variables can be expressed by the {following} adjoint identity{:}
	
	\begin{eqnarray}
	\langle \mathcal{G}(\textbf{q}), \textbf{q}^a \rangle = \langle \textbf{q}, \mathcal{G}^a(\textbf{q}^a) \rangle 
	+ \mathcal{B}(\phi)
	\end{eqnarray}
	where the last term, $\mathcal{B}$, represents a residual from the boundary conditions.
	
	\subsection{Adjoint nonlinear compressible boundary region equations (ACBRE)}
	
	In our particular case of a boundary layer flow over a flat or concave surface with wall transpiration, the ACBRE are derived starting with the integral

	\begin{align}\label{opt}
	& \mathcal{L}(\rho,u,v,w,T,p,\mu,k,v_w,\rho^a,u^a,v^a,w^a,T^a,p^a,\mu^a,k^a,s)   
	\nonumber  \\
	&\quad
	 = \frac{\sigma}{2}  \int_{X_0}^{X_1} \int_{\Gamma} \left\|\frac{\partial v_w}{\partial X}\right\|^2 + \|v_w\|^2 d\Gamma dX 
	 - \int_{0}^{X_0} \int_{\Gamma} s v d\Gamma dX 
	 -  \int_{X_0}^{X_1} \int_{\Gamma} s \left( v - v_w \right) d\Gamma dX
	 - \int_{X_1}^{X_t} \int_{\Gamma} s v  d\Gamma dX \nonumber \\
	 &\quad
	+ \frac{\alpha}{2} \int_{X_{s0}}^{X_{s1}} \int_{\Gamma} \| \tau_w - \tau_0 \|^2 d\Gamma dX \nonumber \\
	&\quad - \int_{0}^{X_t} \int_{\Omega} \rho^a 
	\left(
	\textbf{V} \cdot \nabla \rho
	+ \rho \nabla \textbf{V}
	\right) d\Omega dX 
	\nonumber \\
	&\quad - \int_{0}^{X_t} \int_{\Omega} u^a 
	\left[
	\rho \textbf{V}.\nabla u
	- \nabla_{c} \cdot \left( \mu \nabla_{c} u \right) 
	\right] d\Omega dX 
	\nonumber \\
	&\quad - \int_{0}^{X_t} \int_{\Omega} v^a 
	\left\lbrace
	\rho \textbf{V} \cdot \nabla v 
	+ G_\Lambda u^2 
	+ \frac{\partial p}{\partial y} 
	- \frac{\partial}{\partial y} \left[ \frac{2}{3} \mu \left( 3\frac{\partial v}{\partial y} - \nabla \cdot \textbf{V} \right) \right] 
	- \frac{\partial}{\partial x} \left( \mu \frac{\partial u}{\partial y} \right) 
	- \frac{\partial}{\partial z} \left[ \mu \left( \frac{\partial v}{\partial z} + \frac{\partial w}{\partial y} \right) \right] 
	\right\rbrace d\Omega dX 
	\nonumber \\
	&\quad - \int_{0}^{X_t} \int_{\Omega} w^a 
	\left\lbrace
	\rho \boldsymbol{V} \cdot \nabla w 
	+ \frac{\partial p}{\partial z} 
	- \frac{\partial}{\partial z} \left[ \frac{2}{3} \mu \left( 3\frac{\partial w}{\partial z} - \nabla \cdot \textbf{V} \right) \right] 
	- \frac{\partial}{\partial x} \left( \mu \frac{\partial u}{\partial z} \right)
	- \frac{\partial}{\partial y} \left[ \mu \left( \frac{\partial v}{\partial z} + \frac{\partial w}{\partial y} \right) \right] 
	\right\rbrace d\Omega dX 
	\nonumber \\
	&\quad - \int_{0}^{X_t} \int_{\Omega} T^a 
	\left\lbrace
	\rho \textbf{V} \cdot \nabla T  
	- \frac{1}{Pr} \nabla_{c} \cdot \left( k \nabla_{c} T \right) 
	- (\gamma - 1) M_{\infty}^2\mu \left[ \left( \frac{\partial u}{\partial y} \right)^2
	+ \left( \frac{\partial u}{\partial z} \right)^2 \right] 
	\right\rbrace d\Omega dX
	\nonumber \\
	&\quad - \int_{0}^{X_t} \int_{\Omega} p^a\left(p-\frac{\rho T}{\gamma M_{\infty}^2} R_{e\lambda}^2 \right) d\Omega dX 
	\nonumber \\
	&\quad - \int_{0}^{X_t} \int_{\Omega} \mu^a(\mu-T^b) d\Omega dX 
	\nonumber \\
	&\quad - \int_{0}^{X_t} \int_{\Omega} k^a(k-\mu) d\Omega dX
    \end{align}

	In the above equations, the control is applied in specified intervals $[X_0,X_1]$ only.
	$\Omega$ is the cross-section domain $[0,\infty] \times [z_1,z_2]$ ranging from the wall ($y=0$) to infinity and from $z_1$ to $z_2$ in the spanwise direction, and $\Gamma$ is the wall boundary line for a given $X$, $\tau_w$ is the wall shear stress, $\tau_0$ is a target shear stress (equal to the value corresponding to the Blasius solution), and $[X_{s0},X_{s1}]$ is the interval where the cost function is defined. The last three integrals in (\ref{opt}) are used to enforce the boundary condition on $v$, which includes the wall transpiration. If we take the directional derivative of the Lagrangian with respect to $\rho$, the result is

	\begin{align}
	&\int_{0}^{X_t} \int_{\Omega} 
	\rho^a 
	\left( 
	\textbf{V} \cdot \nabla \delta \rho 
	+ \delta \rho \nabla \cdot \textbf{V} 
	\right)
	+ \delta\rho \textbf{V} \cdot
	\left( 
	  u^a \nabla u
	+ v^a \nabla v
	+ w^a \nabla w
	+ T^a \nabla T
	\right) +\frac{T R_{e\lambda}^2}{\gamma M_\infty^2}\left(v^a\frac{\partial \delta \rho}{\partial y} + w^a\frac{\partial \delta \rho}{\partial z}\right) \nonumber \\
	&\quad \frac{\delta \rho R_{e\lambda}^2}{\gamma M_\infty^2}\left(v^a\frac{\partial T}{\partial y} + w^a\frac{\partial T}{\partial z}\right)
	- p^a \frac{\delta\rho T}{\gamma M_\infty^2} R_{e\lambda}^2
	d\Omega dX = 0
	\end{align}

	The first adjoint equation is obtained by working the integration by parts in $[0,X_t]$ and in $\Omega$. We assume arbitrary variations of $\delta \rho$ in $[0,X_t] \times \Omega$ and that $\delta \rho|_{\Gamma}=0$ (spanwise variation) and $\delta \rho|_{0}^{X_t}=0$ (streamwise variation).

        \begin{align}\label{adj1}
	        &\textbf{V} \cdot \left(u^a \nabla u 
	        + v^a \nabla v +  w^a \nabla w + T^a \nabla T \right)- p^a \frac{T R_{e\lambda}^2}{\gamma M_\infty^2} - \textbf{V} \cdot \nabla  \rho^a - \frac{R_{e\lambda}^2}{\gamma M_\infty^2} T \nabla_c \cdot \textbf{V}^a = 0 \hspace{3mm} on \hspace{3mm} [0,X_t] \times \Omega 
        \end{align}

where $\textbf{V}^{a} = u^{a} \vec{i} + v^{a} \vec{j} + w^{a}\vec{k}$.
	
We obtain the ACBRE corresponding to the rest of the state variables in a similar fashion by taking the directional derivative of the Lagrangian with respect $u$, $v$, $w$, and $T$, respectively as 

\begin{align}\label{adj2}
		&\rho \left(\textbf{V}^{a} \cdot \frac{\partial \textbf{V}}{\partial x} + T^a\frac{\partial T}{\partial x}\right)
		+ 2G_\Lambda u v^a  
		- \rho \frac{\partial \rho^a}{\partial x}
		- \rho \textbf{V} \cdot \nabla u^a 
		- \left(\mu \nabla_{c} \cdot (\nabla_{c} u^a) 
		+ \nabla_{c} \mu \cdot \nabla_{c} u^a\right)
		+ \frac{2}{3} \frac{\partial \mu}{\partial x} (\nabla_{c} \cdot \textbf{V}^{a}) \nonumber \\ 
		&\quad
	    - \left(\frac{1}{3} \mu \nabla_{c} \cdot \frac{\partial \textbf{V}^{a}}{\partial x}
	    + \nabla_{c} \mu \cdot \frac{\partial \textbf{V}^{a}}{\partial x} \right) + 2(\gamma -1)\left(T^a\mu \nabla_c \cdot \nabla_c u + \mu \nabla_c u \cdot \nabla_c T^a + T \nabla_c u \cdot \nabla_c \mu \right) = 0  \hspace{3mm} on \hspace{3mm} [0,X_t] \times \Omega 
\end{align}
\begin{align}\label{adj3}
	&\rho \left(\textbf{V}^{a} \cdot \frac{\partial \textbf{V}}{\partial y} 
	+ T^a\frac{\partial T}{\partial y}\right)
	- \rho \frac{\partial \rho^a}{\partial y} 
	- \rho \boldsymbol{V} \cdot \nabla v^a
	- \Bigg[
	\frac{4}{3}\left(\frac{\partial \mu}{\partial y} \frac{\partial v^a}{\partial y} 
	+ \mu \frac{\partial^2 v^a}{\partial y^2}\right) 
	+ \frac{\partial \mu}{\partial z} \frac{\partial v^a}{\partial z} 
	+ \mu \frac{\partial^2 v^a}{\partial z^2} 
	+ \frac{1}{3} \mu \frac{\partial^2 w^a}{\partial y \partial z} \nonumber \\
	&\quad
	+ \frac{\partial \mu}{\partial z} \frac{\partial w^a}{\partial y}
	\Bigg] 
	+ \frac{2}{3} \frac{\partial \mu}{\partial y} \frac{\partial w^a}{\partial z}  = 0  \hspace{3mm} on \hspace{3mm} [0,X_t] \times \Omega 
\end{align}
\begin{align}\label{adj4}
	&\rho \left(\textbf{V}^{a} \cdot \frac{\partial \textbf{V}}{\partial z} 
	+ T^a\frac{\partial T}{\partial z}\right)
	- \rho \frac{\partial \rho^a}{\partial z} 
	- \rho \boldsymbol{V} \cdot \nabla w^a 
	- \Bigg[
	\frac{4}{3}\left(\frac{\partial \mu}{\partial z} \frac{\partial w^a}{\partial z} 
	+ \mu \frac{\partial^2 w^a}{\partial z^2}\right) 
	+ \frac{\partial \mu}{\partial y} \frac{\partial w^a}{\partial y} 
	+ \mu \frac{\partial^2 w^a}{\partial y^2} 
	+ \frac{1}{3} \mu \frac{\partial^2 v^a}{\partial y \partial z} \nonumber \\
	&\quad
	+ \frac{\partial \mu}{\partial y} \frac{\partial v^a}{\partial z}
	\Bigg] 
	+ \frac{2}{3} \frac{\partial \mu}{\partial z} \frac{\partial v^a}{\partial y}  = 0  \hspace{3mm} on \hspace{3mm} [0,X_t] \times \Omega 
\end{align}
\begin{align}\label{adj5}
	&\rho \boldsymbol{V} \cdot \nabla T^a 
	+ \frac{1}{Pr}\left( \nabla_c k \cdot \nabla_c T^a 
	+ k \nabla_c^2 T^a\right)
	+ p^a \frac{\rho R_{e\lambda}^2}{\gamma M_\infty^2}
	+ \mu^a b T^{b-1}= 0  \hspace{3mm} on \hspace{3mm} [0,X_t] \times \Omega 
\end{align}

The initial and boundary conditions associated with the ACBRE are
	\begin{equation}\label{bc1}
	(\rho^a,u^a,v^a,w^a,T^a)|_{X=X_t} = (0,0,0,0,0) \hspace{2mm} in  \hspace{2mm} \Omega,
	\end{equation}
	\begin{equation}\label{bc2}
	(u^a,v^a,w^a)|_{\Gamma} = 
	\begin{cases}
	(\alpha(\tau_w - \tau_0),0,0) \hspace{2mm} for  \hspace{2mm} X \in [X_{s0},X_{s1}] &  \\
	(0,0,0) \hspace{2mm} otherwise  & 
	\end{cases}
	\end{equation}
	and 
	\begin{equation}\label{bc3}
	(\rho^a,u^a,v^a,w^a,T^a)|_{Y \rightarrow \infty} = (0,0,0,0) \hspace{2mm}
	\end{equation}
	where $\alpha$ is a constant pre-factor that controls the penalization of the wall shear stress. With the state variables $(\rho,u,v,w,T)$ determined from equations (\ref{neq1})-(\ref{neq5}), the ACBRE (\ref{adj1})-(\ref{adj5}) are linear and parabolic, and can be solved via a marching procedure in the backward direction, starting from the terminal streamwise location, $X_t$, towards the initial streamwise location $X_0$. In the present study, the values of the constant factors $\alpha$ and $\sigma$ are set to $1$ and $0.1$, respectively. 
	
	The non time-dependent adjoint variables $p^a$, $\mu^a$, $k^a$ are determined in a similar fashion as
	
\begin{equation}\label{adj6}
	p^a = \nabla_c \cdot \textbf{V}^a
\end{equation}

\begin{equation}\label{adj7}
	k^a = -\frac{1}{Pr}\left( \nabla_c T^a \cdot \nabla_c T\right)
\end{equation}

\begin{align}\label{adj8}
	&\mu^a = k^a + T^a(\gamma -1)M_{\infty}^2\left[\left(\frac{\partial u}{\partial y}\right)^2 + \left(\frac{\partial u}{\partial z}\right)^2\right]
	-\Bigg[
	\nabla_c u^a \cdot \nabla_c u 
	+ \frac{2}{3} \frac{\partial v^a}{\partial y}
	\left(
	3 \frac{\partial v}{\partial y}
	- \nabla \cdot \textbf{V}
	\right)
	+ \frac{2}{3} \frac{\partial w^a}{\partial z}
	\left(
	3 \frac{\partial w}{\partial z}
	- \nabla \cdot \textbf{V}
	\right) \nonumber\\ 
	&\quad + \nabla_c u \cdot \frac{\partial \boldsymbol{V}^a}{\partial x}
	+ \left(
	\frac{\partial v^a}{\partial z} + \frac{\partial w^a}{\partial y}
	\right)
	\left(
	\frac{\partial v}{\partial z} + \frac{\partial w}{\partial y}
	\right)
	\Bigg]
\end{align}

The control variables are updated using the optimality conditions given by 
	
	\begin{equation}\label{opt_cond}
	\sigma\left(v_w - \frac{\partial^2 v_w}{\partial x^2}\right) + s = 0
	\end{equation}
	where $s$ is the Lagrange multiplier corresponding to the transpiration condition. It is obtained from the derivation of the third adjoint equation \eqref{adj3} as result of the integration by parts 
	\begin{equation}
	s = -\rho^a\rho - v^a\rho(u+v+w) + 
	\mu\left( \frac{4}{3} \frac{\partial v^a}{\partial y} + \frac{\partial v^a}{\partial z} - \frac{2}{3} \frac{\partial w^a}{\partial z} + \frac{\partial w^a}{\partial y}\right)
	\end{equation}

The control algorithm starts with the solution to the NCBRE for the uncontrolled boundary layer, followed by the solution to the ACBRE (note that the ACBRE depend on the solution to the NCBRE). The difference between the wall shear stress and the original laminar wall shear stress is then compared to a desired value; if the difference is larger than a given threshold then the steepest descent method is utilized to determine the new wall transpiration velocity $v_w$. Once these are determined, the loop is repeated. In total, we apply 10 control iterations.  

	
The adjoint equations (\ref{adj1})-(\ref{adj5}) and the associated initial and boundary conditions (\ref{bc1})-(\ref{bc3}) are solved numerically on the same grid as the original NCBRE state equations (\ref{neq1})-(\ref{neq5}), and utilizing the same numerical algorithm (as for the NCBRE equations), except that the marching is preformed backwards, starting from the terminal streamwise location.

Thanks to the robust mathematical model of the NCBRE and the associated numerical algorithm investigated in our previous work, Es-Sahli \textit{et al.}\cite{Es-Sahli2}, the overall computational cost of the control algorithm is reduced.

\subsection{Flow domain and numerical simulation setup}

The 3D flow domain including region III (starting at the initial streamwise location $X_0$) extends $50\times\lambda^*$ in the streamwise direction, $10\times\lambda^*$ in the wall-normal direction, and $1\times\lambda^*$ in the spanwise direction.

In terms of the numerical setup of the simulations, we use second and fourth-order finite-difference schemes to discretize the spatial derivatives in the wall-normal and spanwise directions, respectively. We implement periodic conditions in the spanwise direction to avoid compromising the stability and a staggered arrangement in the wall-normal direction to avoid decoupling between the velocity and pressure (this in turn visually extends the domain to $2\times\lambda^*$ in the spanwise direction, as seen in subsequent figures in the results section). We apply a first-order finite-difference marching scheme in the streamwise direction and converge the set of equations using a nonlinear time relaxation method.

\begin{figure}[hbt!]
 \begin{center}
 \includegraphics[width=\linewidth]{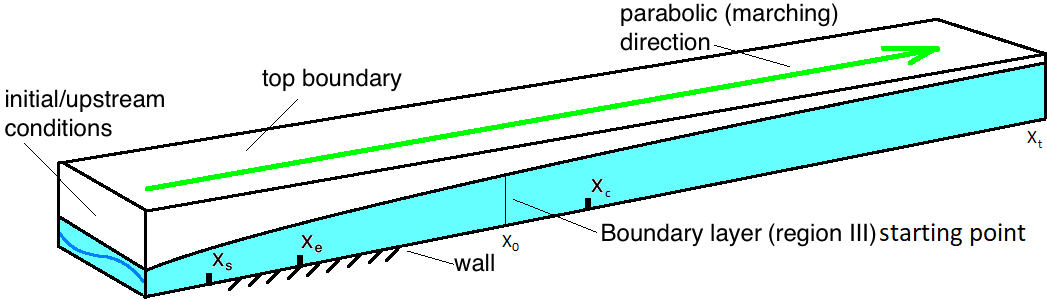}
 \end{center}
 \caption{3D flow domain}
 \label{f2}
\end{figure}

Appropriate initial/upstream and boundary conditions are necessary to close the problem. In this work, we trigger the perturbations in the boundary layer using a small amplitude transpiration velocity ($v_t$) at the wall, in the form;
\begin{align}\label{dist}
v_t = A \sin \left[ \pi \frac{(x-x_s)}{(x_e-x_s)} \right]^2 \cos \left( \pi \frac{z}{\lambda} \right)
\end{align}
where $A$ is the amplitude, $x_s=1.5\lambda^*$ and $x_e=4.5\lambda^*$ are the start and end streamwise locations of the perturbation, respectively. $X_0$ (set to 0.8m from the leading edge of the boundary layer) denotes the initial streamwise location where region III starts. We denote the streamwise location of the control starting point (measured from $X_0$) where the transpiration velocity control kicks off by $X_c$. Figure \ref{f2} illustrates the different components of the domain and their respective placement. 

We numerically solve the NCBRE using an algorithm similar to the one employed by Sescu \& Thompson \cite{Sescu3} in the incompressible regime. In the compressible regime, the continuity equation does not re{sult in} the {usual} divergence-free condition, {thus} making the {compressible} set of equations {numerically} stiff. Instead,  it is an equation for density. Therefore, the same numerical algorithm is much faster when applied to the compressible regime. We impose vanishing gradients for all dependent variables along the top boundary.

We generate the mean inflow condition from a similarity solution obtained using the Dorodnitsyn-Howarth coordinate transformation
$
\bar{Y}(x,y) = \int_{0}^{y} \rho(x,\tilde{y}) d \tilde{y}
$.
We define the similarity variable as
$
\eta = \bar{Y}\left( Re_{x}/2x \right)^{1/2},
$
where $Re_{x}$ is the Reynolds number calculated based on the freestream velocity and the distance from the leading edge $X_0$. 

We express the base velocity and temperature as
\begin{equation}\label{ssd}
U = F'(\eta), \hspace{2mm} V = (2 x Re_{x})^{-1/2} (\eta_c T F' - T F), \hspace{2mm} T = T(\eta)
\end{equation}
where the prime superscript represents differentiation with respect to $\eta$, and $\eta_c = 1/T \int_{0}^{\eta} T(\tilde{\eta}) d \tilde{\eta}$. $F$ and $T$ satisfy the following coupled equations
\begin{align}\label{bl}
& \left( \frac{\mu}{T} F'' \right)' + F F'' = 0, \nonumber \\
&\quad \frac{1}{Pr} \left( \frac{\mu}{T} T' \right)' + F T' + (\gamma - 1) M^2 \frac{\mu}{T} F''^2 = 0,
\end{align}
subject to the boundary conditions
$
F(0)=F'(0)=0, \hspace{4mm} T'(0)=0, \hspace{4mm} F' \rightarrow 1, \hspace{4mm} T \rightarrow 1 \rightarrow
\hspace{2mm} as \hspace{2mm}  \eta \rightarrow \infty
$. Equations (\ref{bl}) were solved numerically to determine $F$ and $T$, which were then used in equations (\ref{ssd}) to obtain the mean inflow condition. 

\section{Results}

We present the numerical simulation results of the boundary layer optimal control approach. We assess the efficiency of the control approach across different flow conditions, including two high-supersonic cases ($M_\infty = 3$ and $M_\infty = 4$) and two hypersonic cases ($M_\infty = 5$ and $M_\infty = 6$), by examining two main criterion; the spanwise averaged wall shear stress and the vortex kinetic energy. Since we are using wall transpiration to control the flow, the cumulative blowing or suction at the wall is employed such that a zero mass flow rate is maintained to avoid injecting or absorbing mass into/from the flow.

We evaluate the spanwise averaged wall shear stress using the integral
\begin{equation}\label{wall shear stress}
\tau_w(x) =  \frac{1}{(z_2 - z_1)}\intop_{z_1}^{z_2}  \left.  \frac{\partial u}{\partial y} \right|_{y=0} (x,0,z)  dz
\end{equation} 
While we calculate the vortex kinetic energy distribution according to the characteristic length $L$ as 
\begin{align}\label{jj}
&E(x) = \intop_{z_1}^{z_2}  \intop_{-\infty}^{\infty} \left| u(x,y,z) - u_m(x,y) \right|^{2} +  \left| v(x,y,z) - v_m(x,y) \right|^{2} \\ \nonumber
&\quad +  \left| w(x,y,z) - w_m(x,y) \right|^{2} dzdy
\end{align}
where $u_m(x,y)$, $v_m(x,y)$, and $w_m(x,y)$ are the spanwise mean components of velocity, and $z_1$ and $z_2$ are the coordinates of the boundaries in the spanwise direction. 

\begin{figure}[H]
 \begin{center}
   \includegraphics[width=5cm]{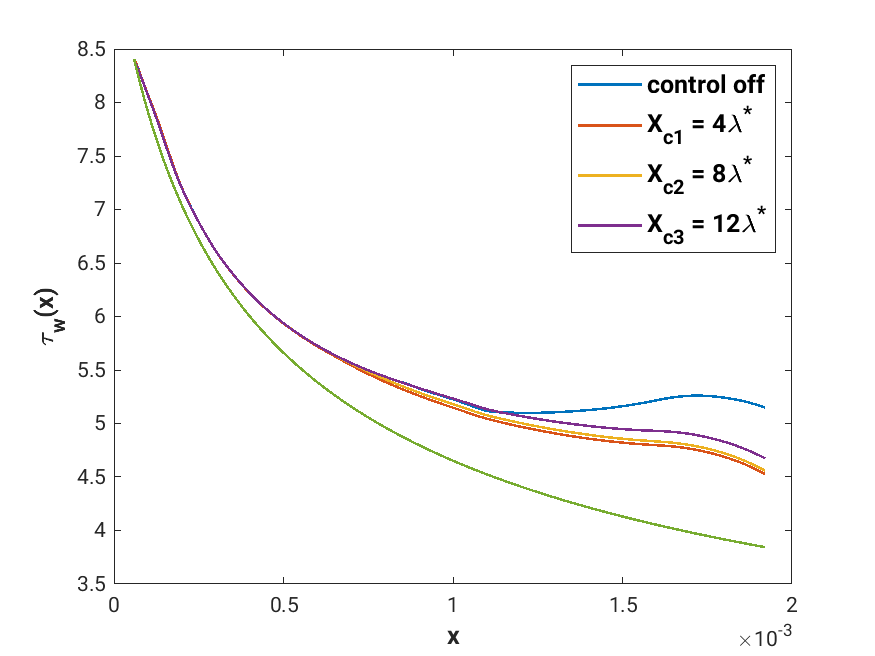}
   \includegraphics[width=5cm]{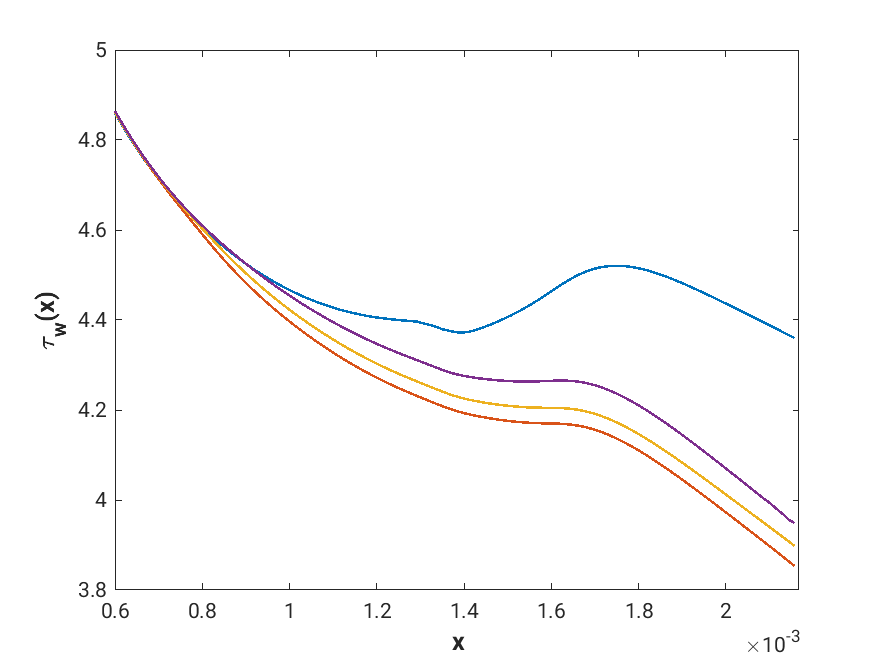} \\
   \hspace{2mm} (a) \hspace{4.3cm} (b)\\
   \includegraphics[width=5cm]{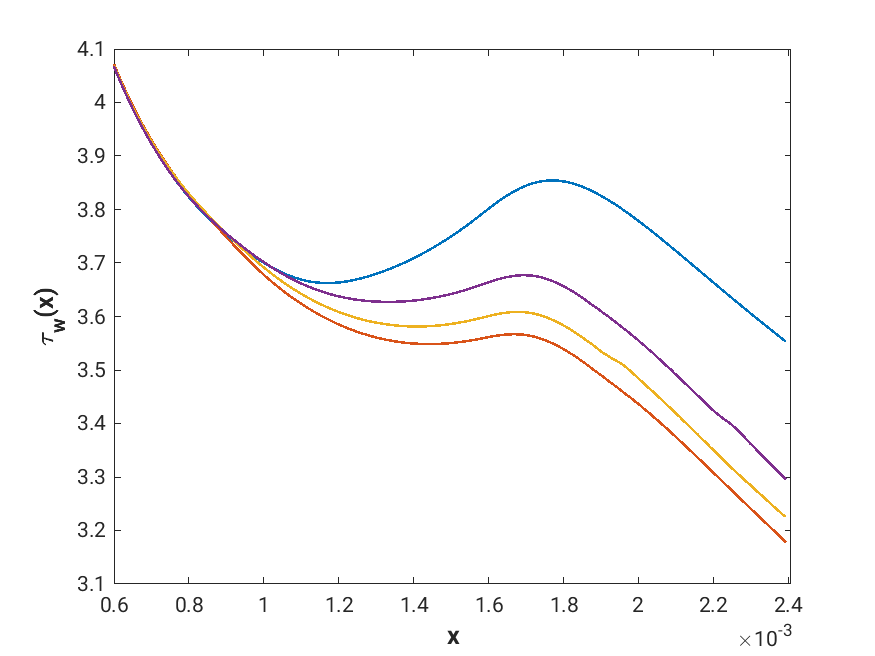} 
   \includegraphics[width=5cm]{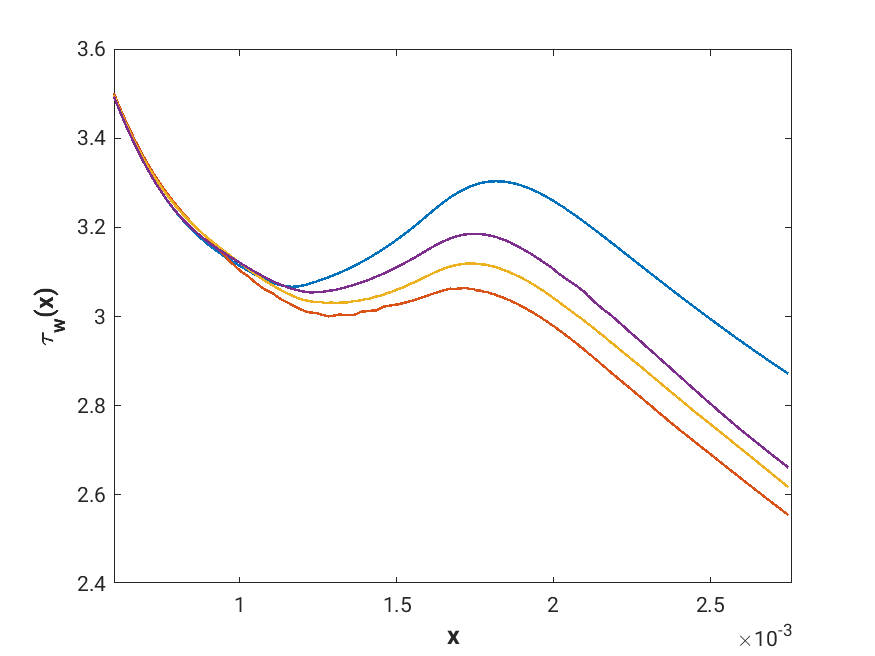} \\
   \hspace{2mm} (c) \hspace{4.3cm} (d)
 \end{center}
  \caption{Effect of the optimal control starting point on the wall shear stress for $M_\infty = 3 \hspace{1mm}(a)$, $M_\infty = 4 \hspace{1mm}(b)$, $M_\infty = 5 \hspace{1mm}(c)$, and $M_\infty = 6 \hspace{1mm}(d)$}
  \label{f3}
\end{figure}

\begin{figure}[H]
 \begin{center}
   \includegraphics[width=5cm]{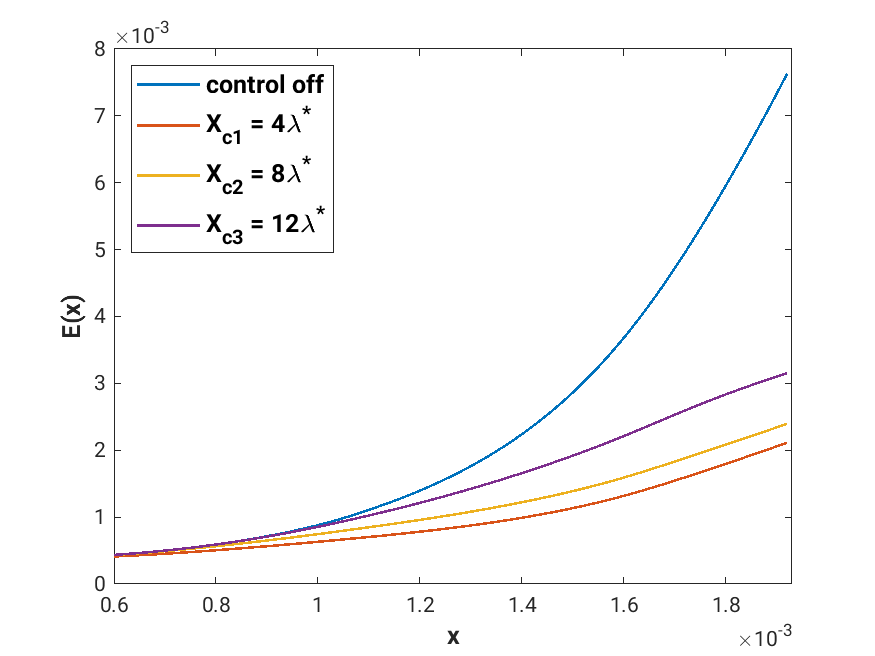}
   \includegraphics[width=5cm]{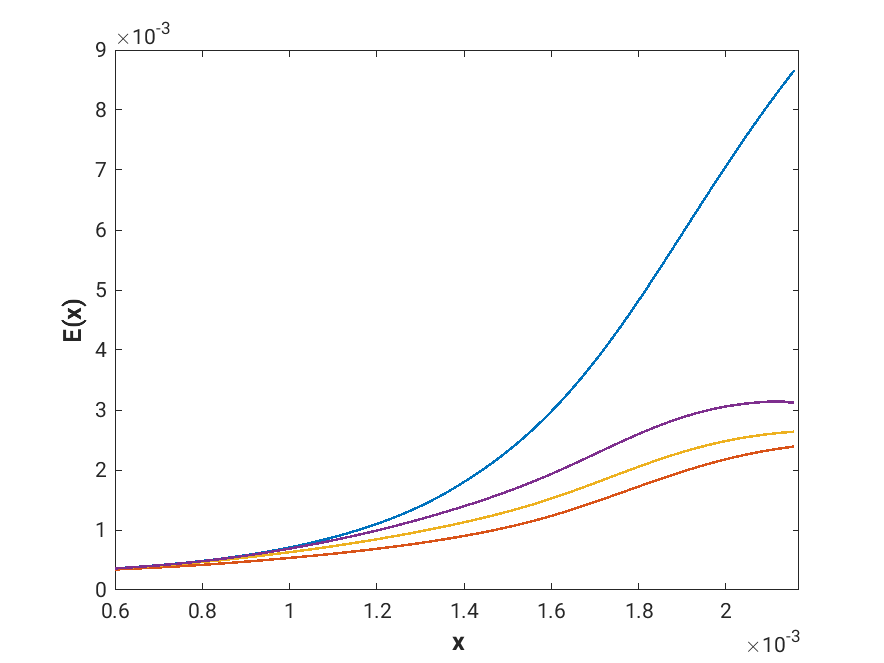} \\
   \hspace{2mm} (a) \hspace{4.3cm} (b)\\
   \includegraphics[width=5cm]{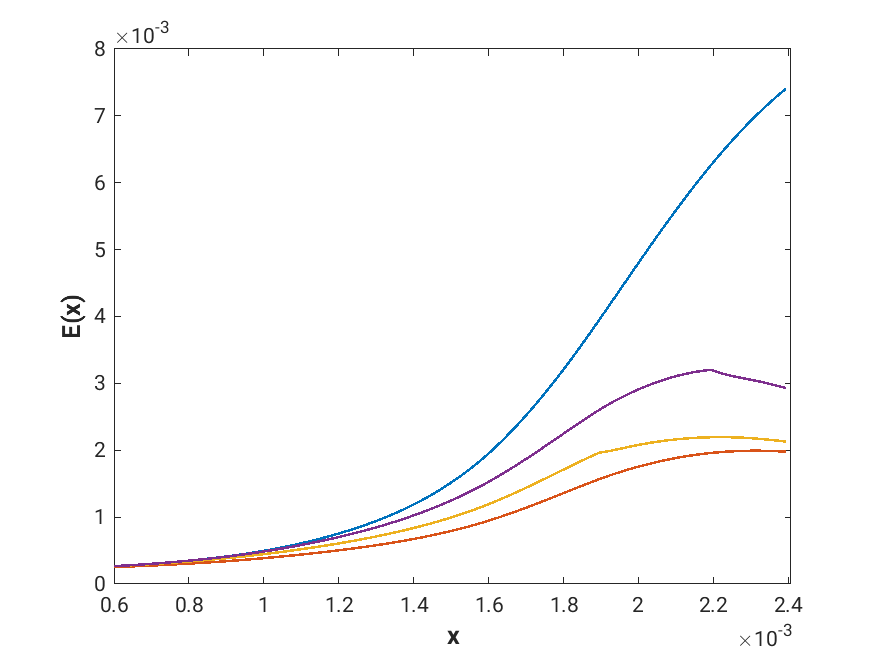}
   \includegraphics[width=5cm]{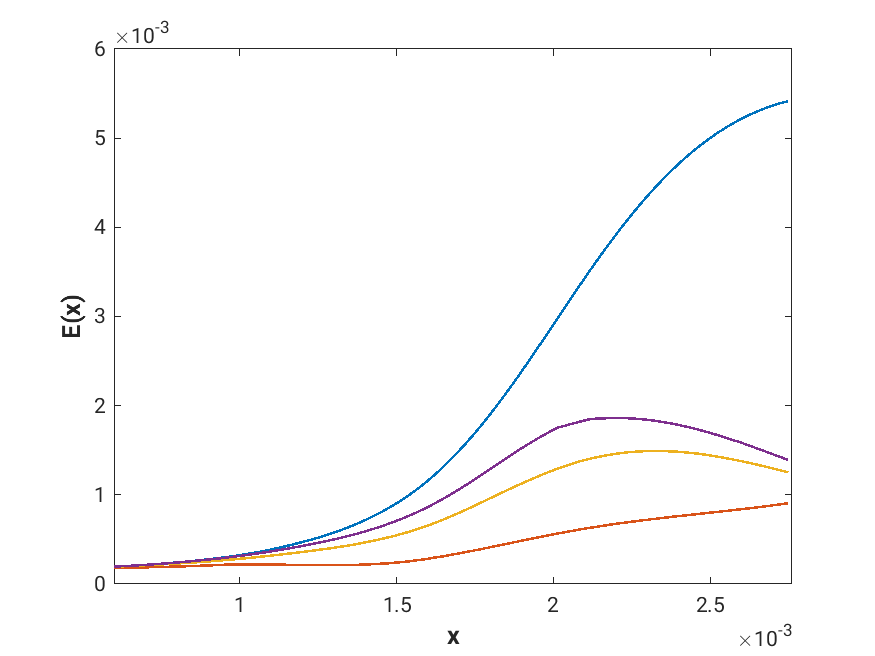} \\
   \hspace{2mm} (c) \hspace{4.3cm} (d)
 \end{center}
  \caption{Effect of the optimal control starting point on the vortex energy for $M_\infty = 3 \hspace{1mm}(a)$, $M_\infty = 4 \hspace{1mm}(b)$, $M_\infty = 5 \hspace{1mm}(c)$, and $M_\infty = 6 \hspace{1mm}(d)$}
  \label{f4}
\end{figure}

We test three different $X_c$ locations to determine the optimal streamwise location to apply the transpiration velocity control; $X_{c_1} = 4\lambda^*$, $X_{c_2} = 8\lambda^*$, and $X_{c_3} = 12\lambda^*$. In figures \ref{f3} and \ref{f4}, we compare the wall shear stress and vortex energy of the uncontrolled flow with that of the last ($10^{th}$) control iteration of each $X_c$ for all considered Mach numbers. We notice that as we move the control streamwise location starting point farther from $X_0$ (as $X_c$ increases), the control approach effect on the wall shear stress and vortex kinetic energy weakens due to the shrinkage of the control surface. Therefore, for all Mach numbers, starting the control mechanism at $X_{c_1}$ results in the highest reduction in the wall shear stress and the vortex kinetic energy of the centrifugal instabilities. We quantitatively assess the drop in the wall shear stress $\tau_w$ and vortex kinetic energy $E$ due to the control method by calculating the decrease percentage at the terminal streamwise location ($X_t$) of the $10^{th}$ control iteration with respect to the uncontrolled boundary layer flow parameters (considering the Blasius solution as a base reference) as 
\begin{equation}\label{decrease persent}
decrease(\%) =  \left|\frac{a(x_{max})-a_u(x_{max})}{a_u(x_{max}) - a_{b}(x_{max})}\right|\times 100 \nonumber
\end{equation} 
where $a$ is a generic variable representing $\tau_{w}$ or $E$, and the subscripts $u$ and $b$ denote the uncontrolled flow and Blasius solution parameters. Results are summarized in table \ref{tab:table1}.
\begin{table}[hbt!]
\caption{\label{tab:table1} Percentage drop in wall shear stress and vortex kinetic energy due to the control approach at the control locations $X_{c_1}$, $X_{c_2}$, and $X_{c_3}$ for all considered Mach numbers.}
\centering
\begin{tabular}{lcccccc}
&\multicolumn{3}{c}{\textit{Wall shear stress} $\tau_w$}&\multicolumn{3}{c}{\textit{Vortex energy} $E$}\\
\textit{Mach number} & $X_{c_1}$ & $X_{c_2}$ & $X_{c_3}$ & $X_{c_1}$ & $X_{c_2}$ & $X_{c_3}$\\
\hline\\
$M_\infty=3$&$47.7$&$45.28$&$36.46$&$72.31$&$68.61$&$58.65$ \\
$M_\infty=4$&$36.64$&$33.41$&$29.77$&$72.38$&$69.52$&$63.98$ \\
$M_\infty=5$&$29.56$&$25.80$&$20.27$&$73.27$&$71.24$&$60.42$ \\
$M_\infty=6$&$28.51$&$22.90$&$18.94$&$83.27$&$76.88$&$74.30$
\end{tabular}
\end{table}

The reduction in the wall shear stress directly translates to a decrease in the boundary layer skin friction. While the reduction in the kinetic energy of the flow delays the occurrence of the secondary instabilities leading to transition, consequently elongating the laminar region of the boundary layer. Ultimately, this results in a more stable and attached boundary layer.

The present findings indicate that a larger control surface results in a larger drop in the wall shear stress and the vortex kinetic energy of the G\"{o}rtler vortices. On the other hand, increasing the control surface adversely affects the computational efficiency of the method as the simulation becomes more computationally expensive, although the increase in the computational cost is relatively insignificant. 

Although the control streamwise locations $X_{c_2}$ and $X_{c_3}$ also cause a significant decrease in the wall shear stress and vortex kinetic energy compared to the uncontrolled case, for what follows, we exclusively show results for the $X_{c_1}$ case. 

\begin{figure}[H]
 \begin{center}
   \includegraphics[width=5cm]{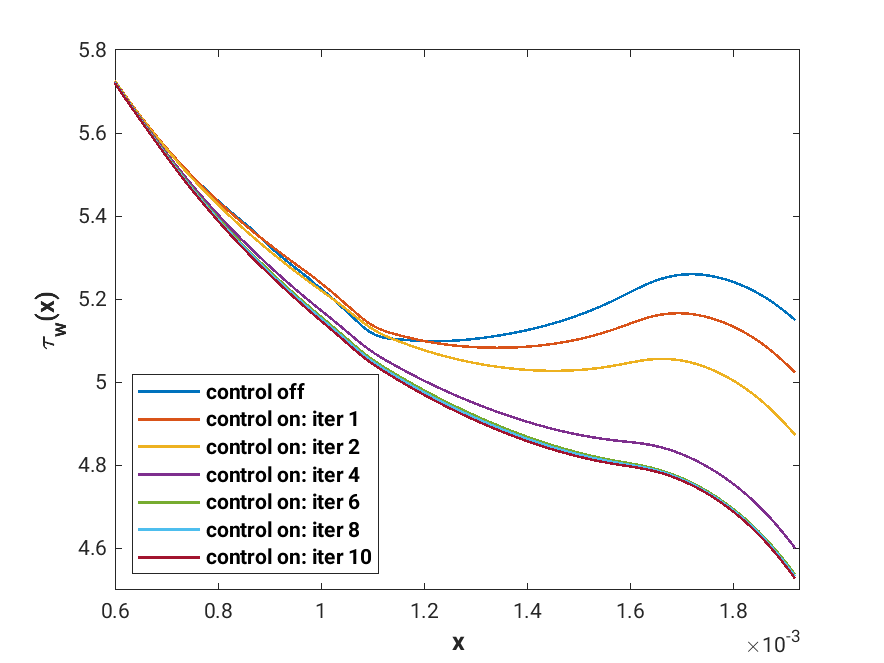}
   \includegraphics[width=5cm]{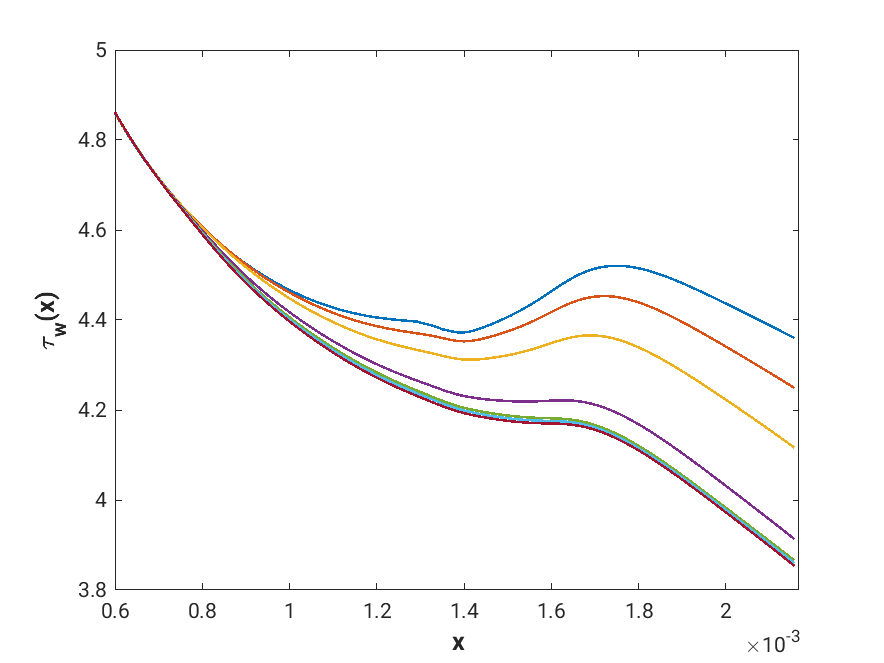} \\
   \hspace{2mm} (a) \hspace{4.3cm} (b)\\
   \includegraphics[width=5cm]{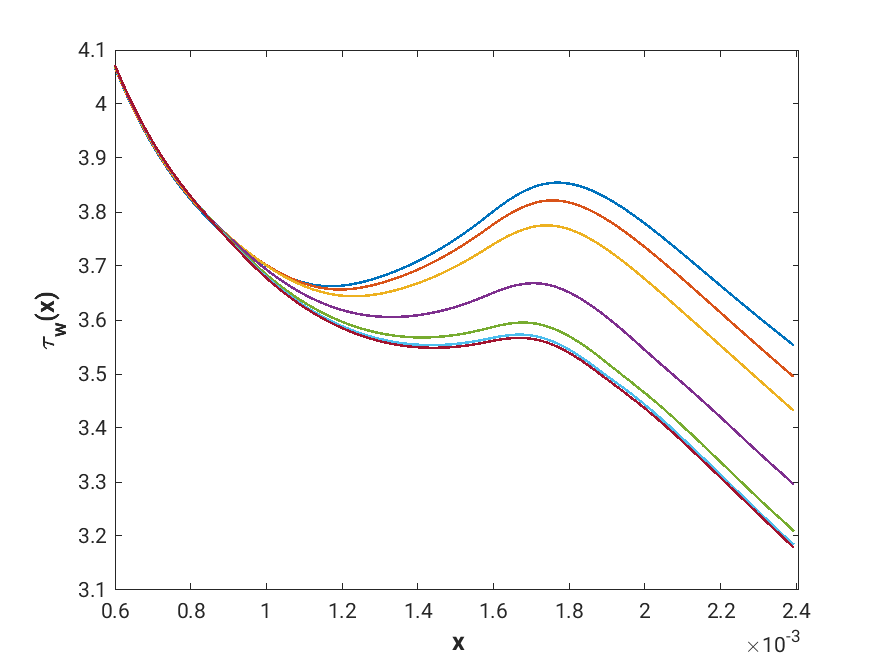}
   \includegraphics[width=5cm]{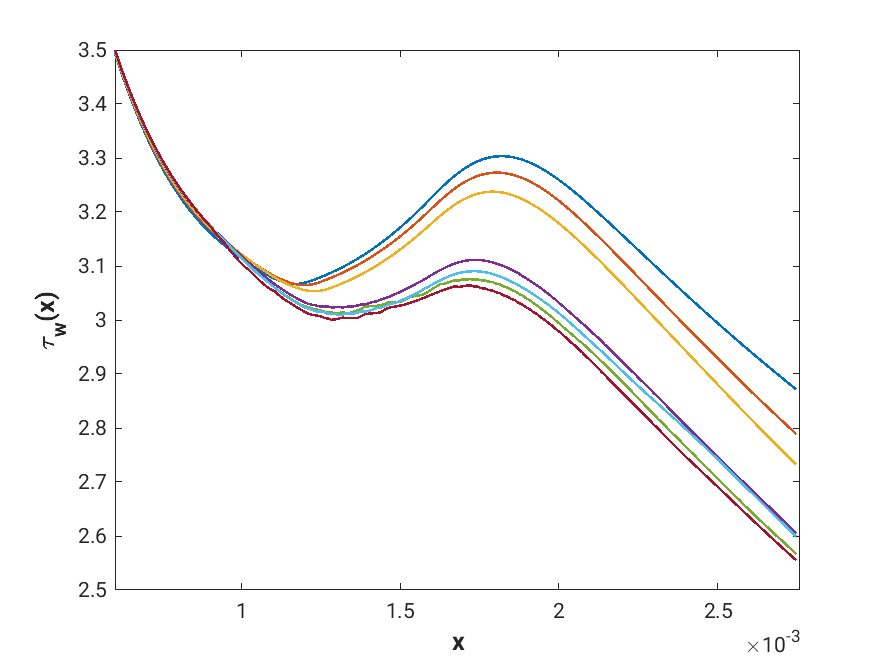} \\
   \hspace{2mm} (c) \hspace{4.3cm} (d)
 \end{center}
  \caption{Effect of increasing optimal control iterations on the wall shear stress for $M_\infty = 3 \hspace{1mm}(a)$, $M_\infty = 4 \hspace{1mm}(b)$, $M_\infty = 5 \hspace{1mm}(c)$, and $M_\infty = 6 \hspace{1mm}(d)$}
  \label{f5}
\end{figure}

\begin{figure}[H]
 \begin{center}
   \includegraphics[width=5cm]{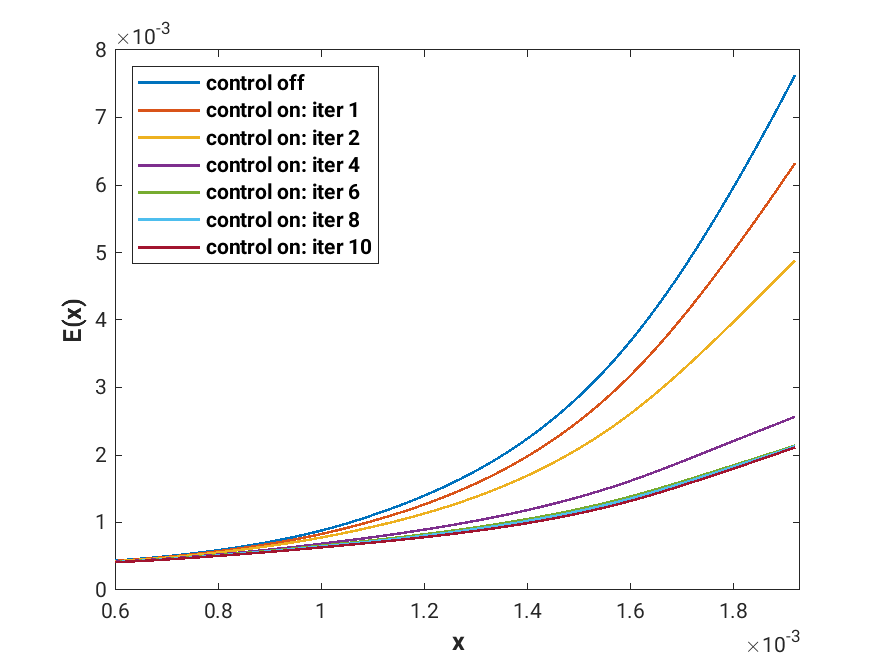}
   \includegraphics[width=5cm]{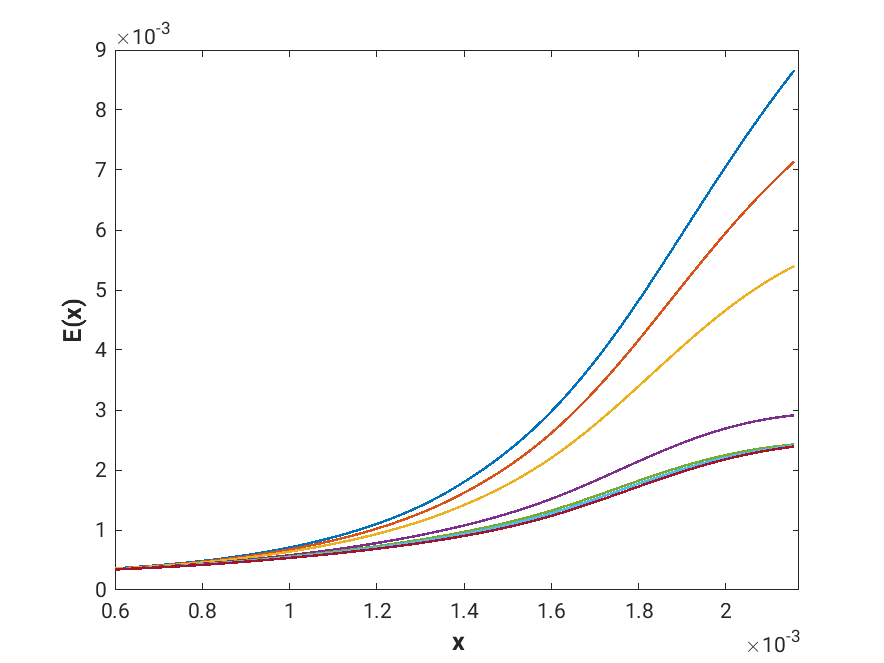} \\
   \hspace{2mm} (a) \hspace{4.3cm} (b)\\
   \includegraphics[width=5cm]{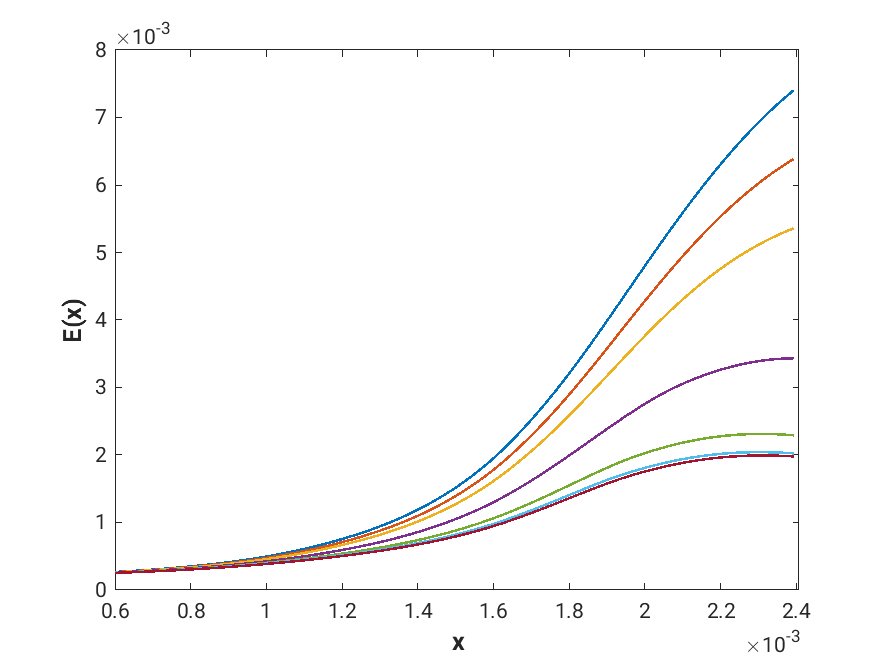}
   \includegraphics[width=5cm]{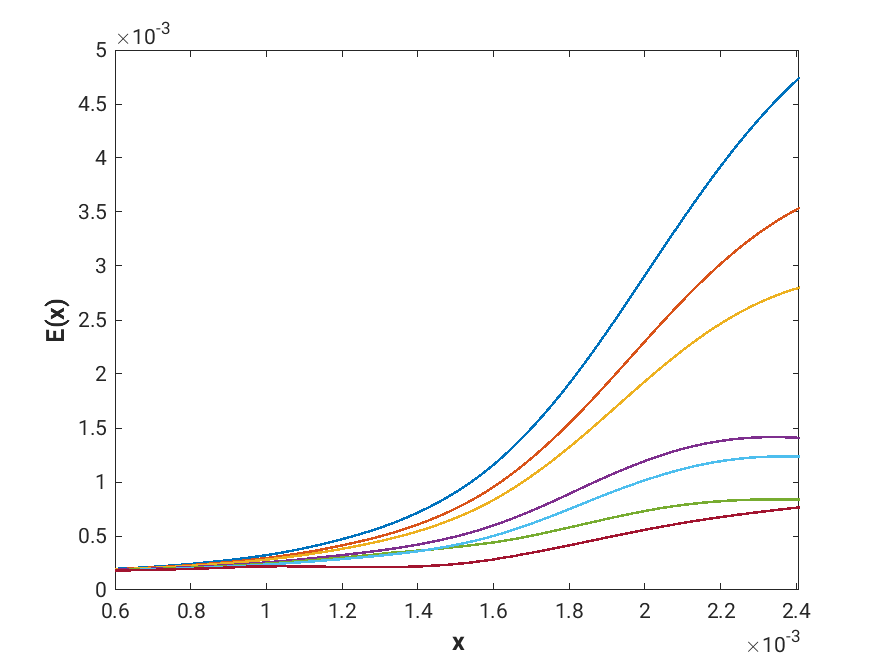} \\
   \hspace{2mm} (c) \hspace{4.3cm} (d)
 \end{center}
  \caption{Effect of increasing optimal control iterations on the vortex energy for $M_\infty = 3 \hspace{1mm}(a)$, $M_\infty = 4 \hspace{1mm}(b)$, $M_\infty = 5 \hspace{1mm}(c)$, and $M_\infty = 6 \hspace{1mm}(d)$}
  \label{f6}
\end{figure}

Next, we consider the effect of increasing the optimal control iterations on the wall shear stress and the vortex kinetic energy, as shown in figures  \ref{f5} and \ref{f6}. As priorly mentioned, we apply a total of ten control iterations. However, we reduce the number of curves and only plot six of the control iterations and the uncontrolled flow case to avoid clustered figures. As expected, more control iterations lower the wall shear stress and vortex kinetic energy of the boundary layer flow. We also observe the number of iterations required to achieve a maximum decrease in the shear stress and vortex kinetic energy scales with the Mach number. For instance, for the two high-supersonic cases, the method achieves maximum reduction after six iterations (figures \ref{f5} and \ref{f6} a and b). On the other hand, the method requires ten iterations for the two hypersonic cases (figures \ref{f5} and \ref{f6} c and d). 

\begin{figure}[H]
 \begin{center}
   \includegraphics[width=4cm]{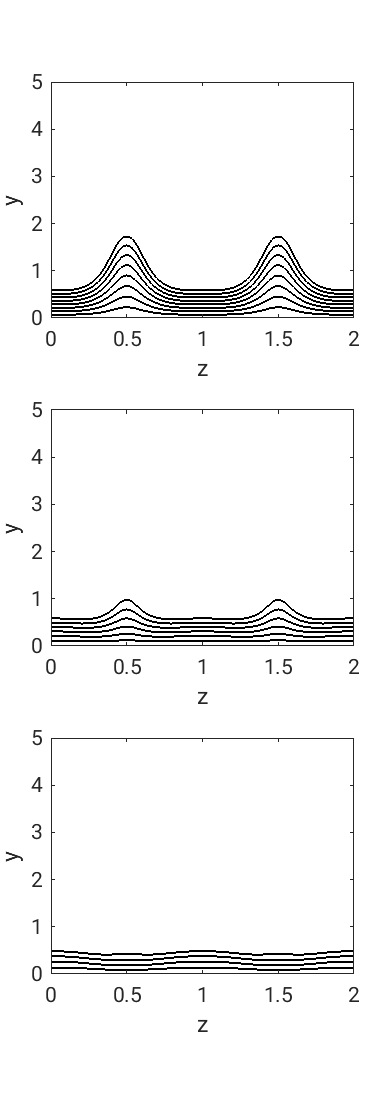}
   \includegraphics[width=4cm]{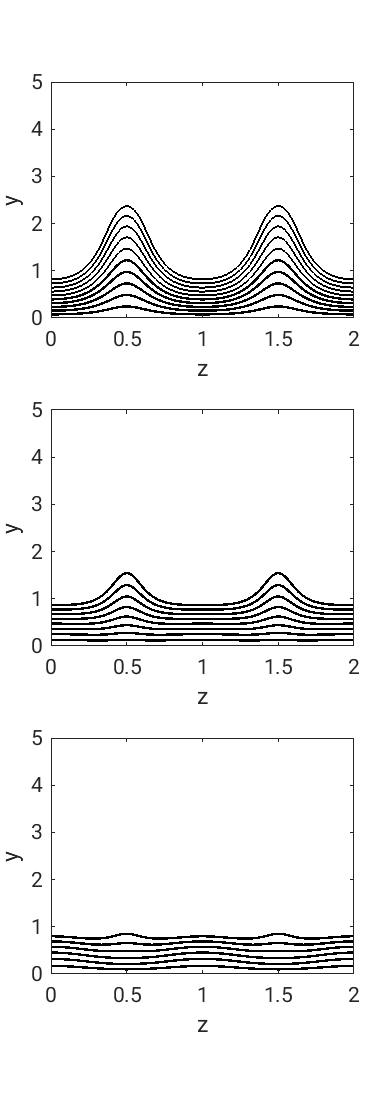}
    \includegraphics[width=4cm]{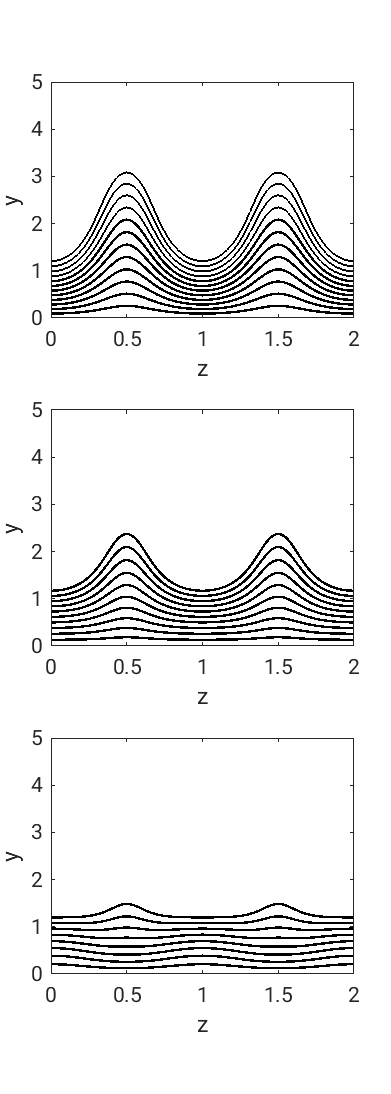}
   \includegraphics[width=4cm]{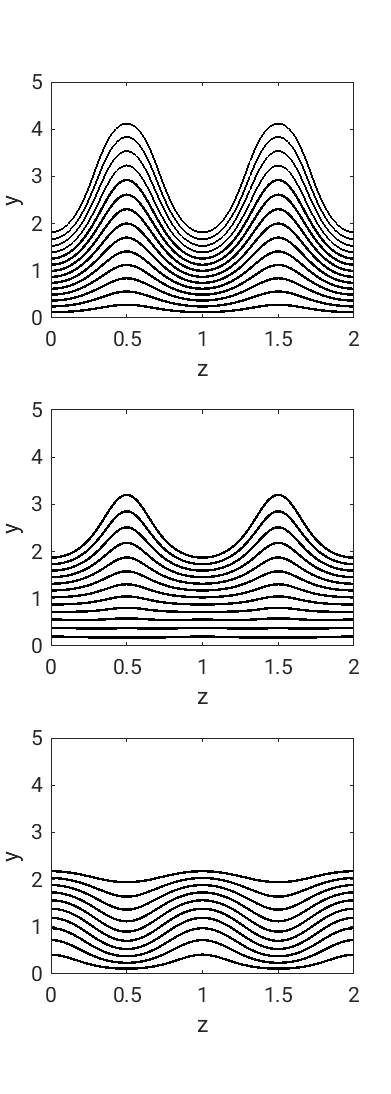}\\
   \hspace{1mm} (a) \hspace{36mm} (b) \hspace{36mm} (c) \hspace{36mm} (d)
 \end{center}
 \caption{Streamwise velocity contours in the YZ-plane for the uncontrolled flow (top), $3^{rd}$ control iteration (middle), and $10^{th}$ control iteration (bottom) for $M_\infty = 3 \hspace{1mm}(a)$, $M_\infty = 4 \hspace{1mm}(b)$, $M_\infty = 5 \hspace{1mm}(c)$, and $M_\infty = 6 \hspace{1mm}(d)$}
  \label{f7}
\end{figure}

The streamwise velocity contours in figure \ref{f7} visually show the effect of the control approach on the centrifugal instabilities of the boundary layer. In the case of the uncontrolled flow (first contour from the top), the G\"{o}rtler vortices correspond to fully developed mushroom-like structures with alternating low and high-speed streaks in the spanwise direction. As we apply more control iterations, the shape of the instabilities, especially in the region closest to the wall, considerably changes. Developed instability structures gradually flatten as the number of control iterations increases due to the drop in the wall shear stress and vortex energy levels in the boundary layer, as discussed above. 

\begin{figure}[H]
 \begin{center}
   \includegraphics[width=8cm]{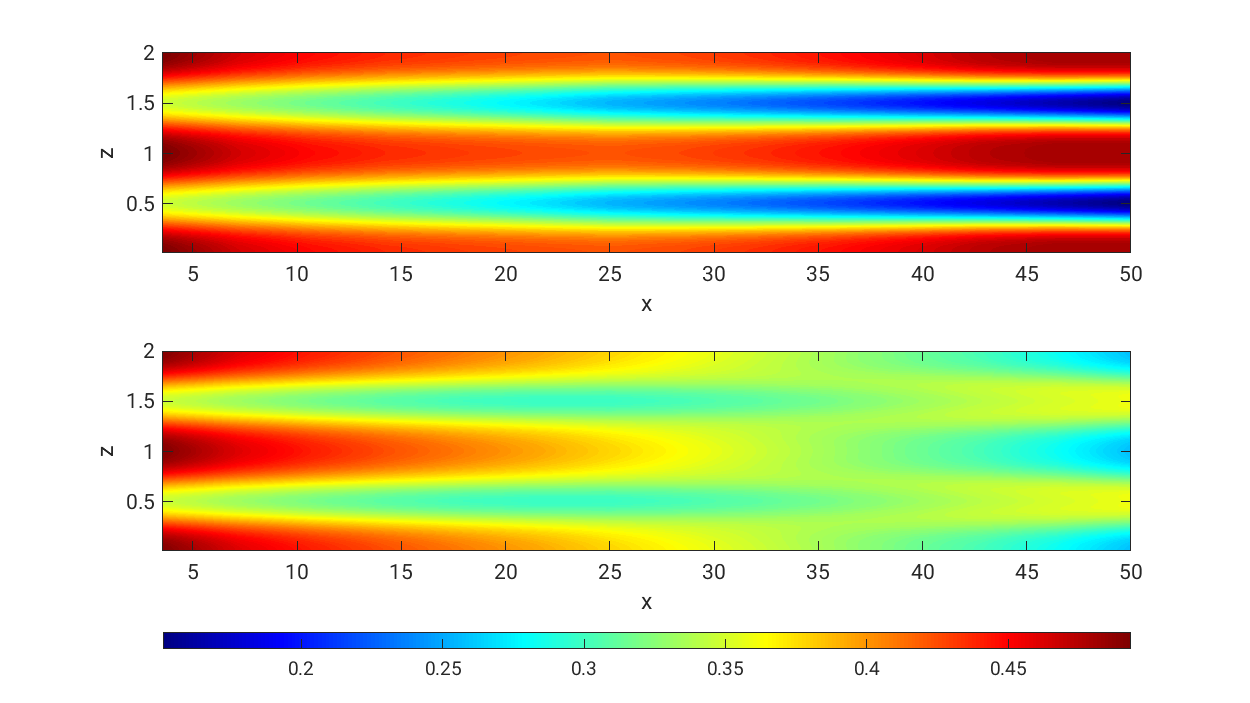}
   \includegraphics[width=8cm]{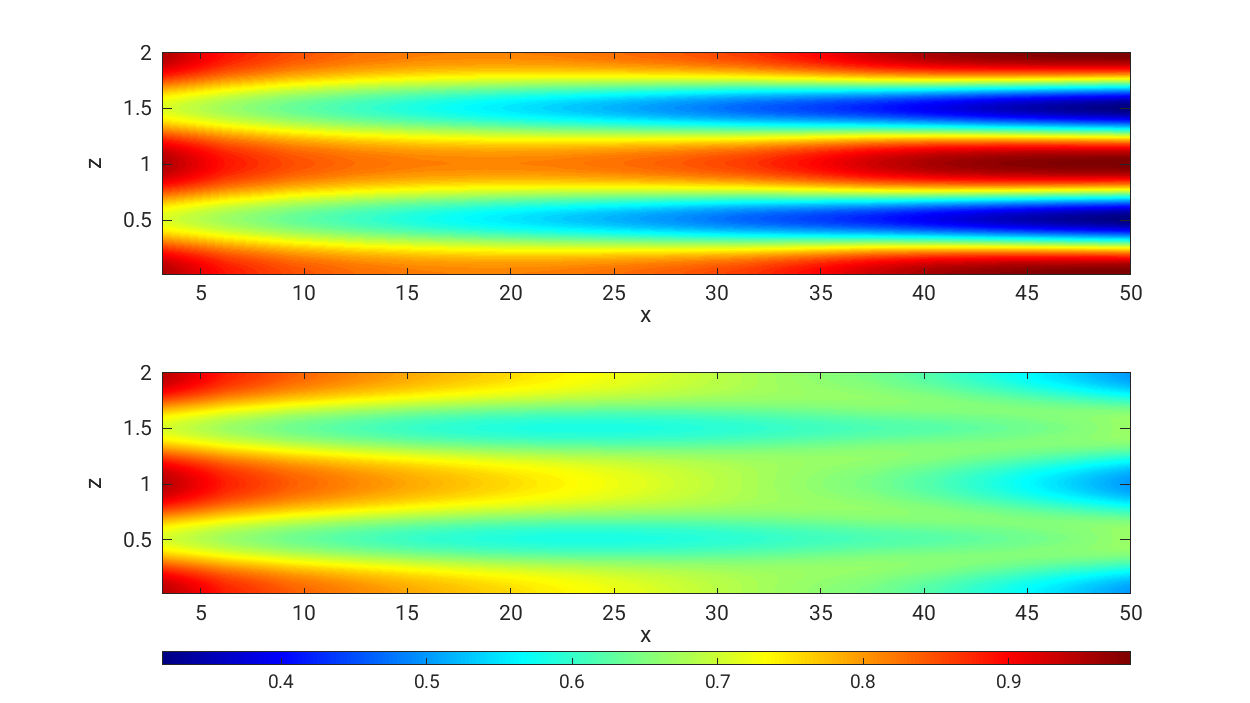}\\
   \hspace{2mm} (a) \hspace{8cm} (b)\\
   \includegraphics[width=8cm]{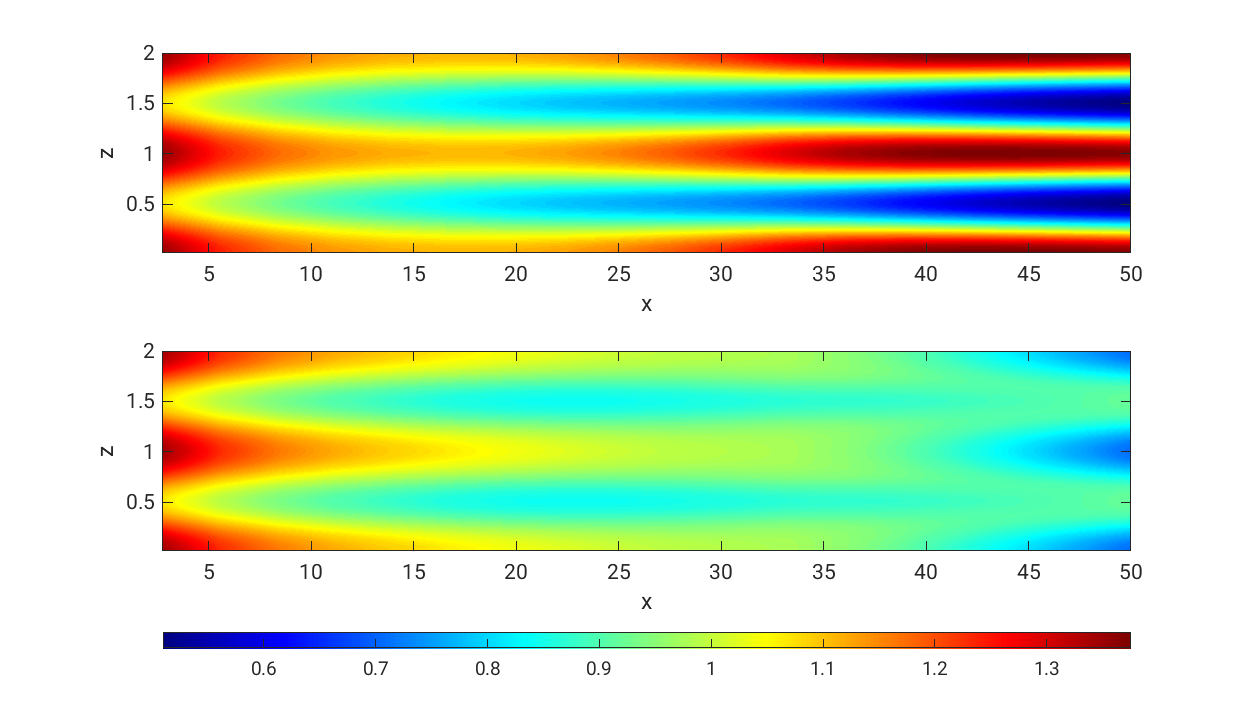}
   \includegraphics[width=8cm]{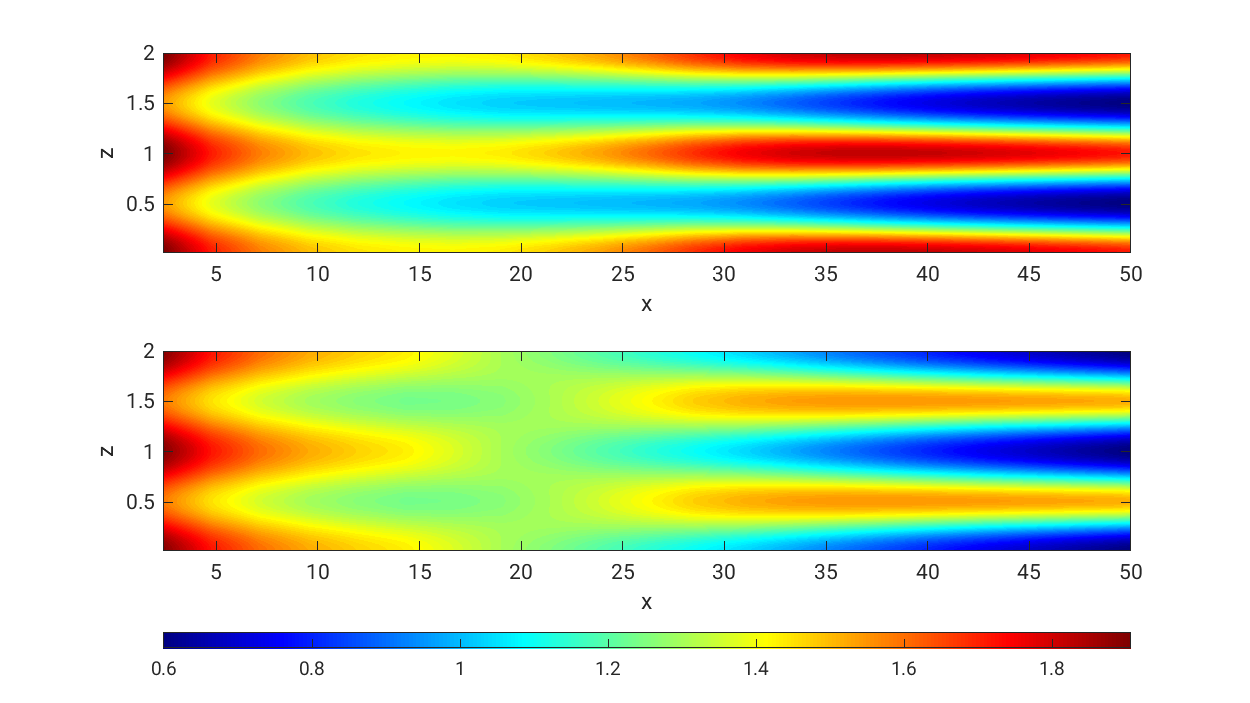}\\
   \hspace{2mm} (c) \hspace{8cm} (d)
 \end{center}
  \caption{Streamwise velocity contours in the ZX-plane for the uncontrolled flow (top) and $10^{th}$ control iteration (bottom) for $M_\infty = 3 \hspace{1mm}(a)$, $M_\infty = 4 \hspace{1mm}(b)$, $M_\infty = 5 \hspace{1mm}(c)$, and $M_\infty = 6 \hspace{1mm}(d)$}
  \label{f8}
\end{figure}

\begin{figure}[H]
 \begin{center}
   \includegraphics[width=8cm]{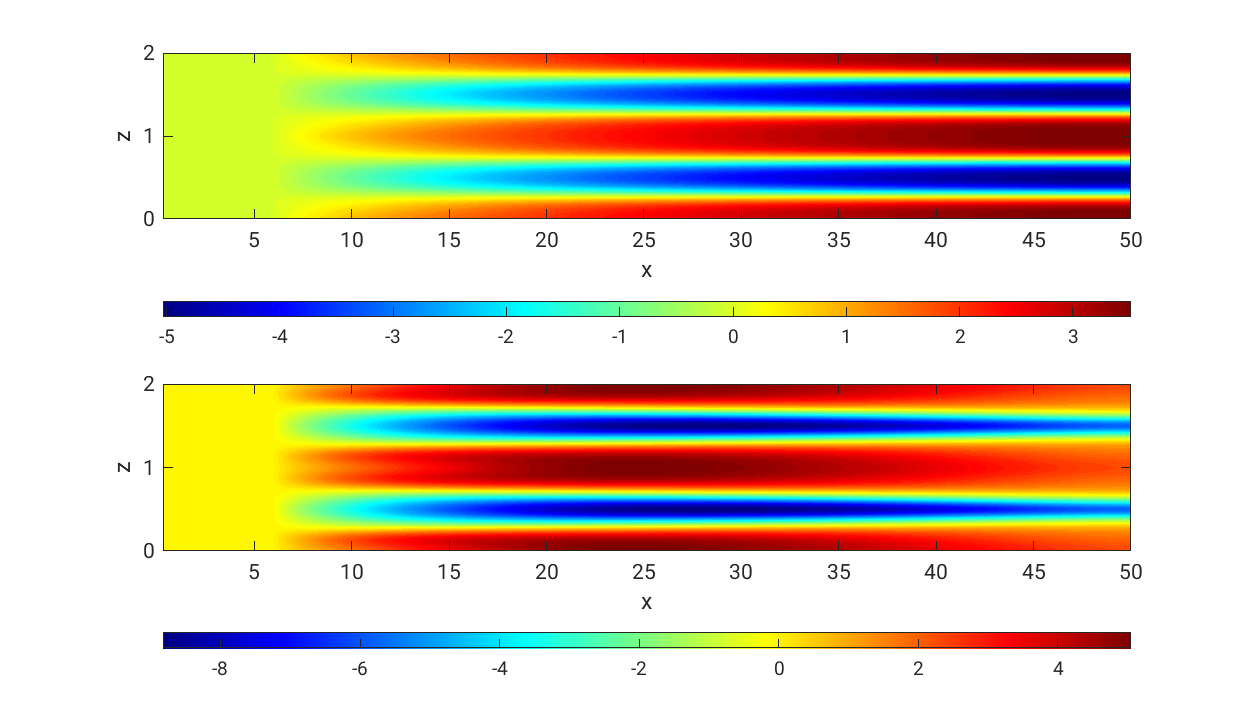}
   \includegraphics[width=8cm]{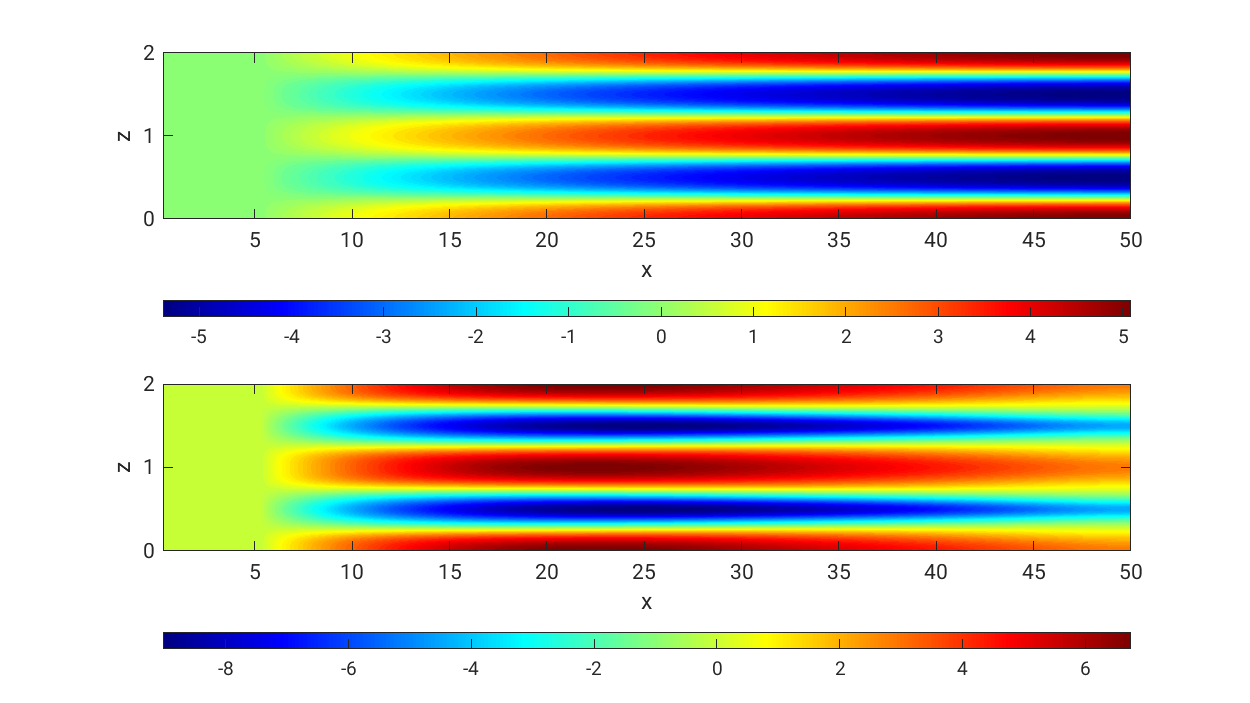}\\
   \hspace{2mm} (a) \hspace{8cm} (b)\\
   \includegraphics[width=8cm]{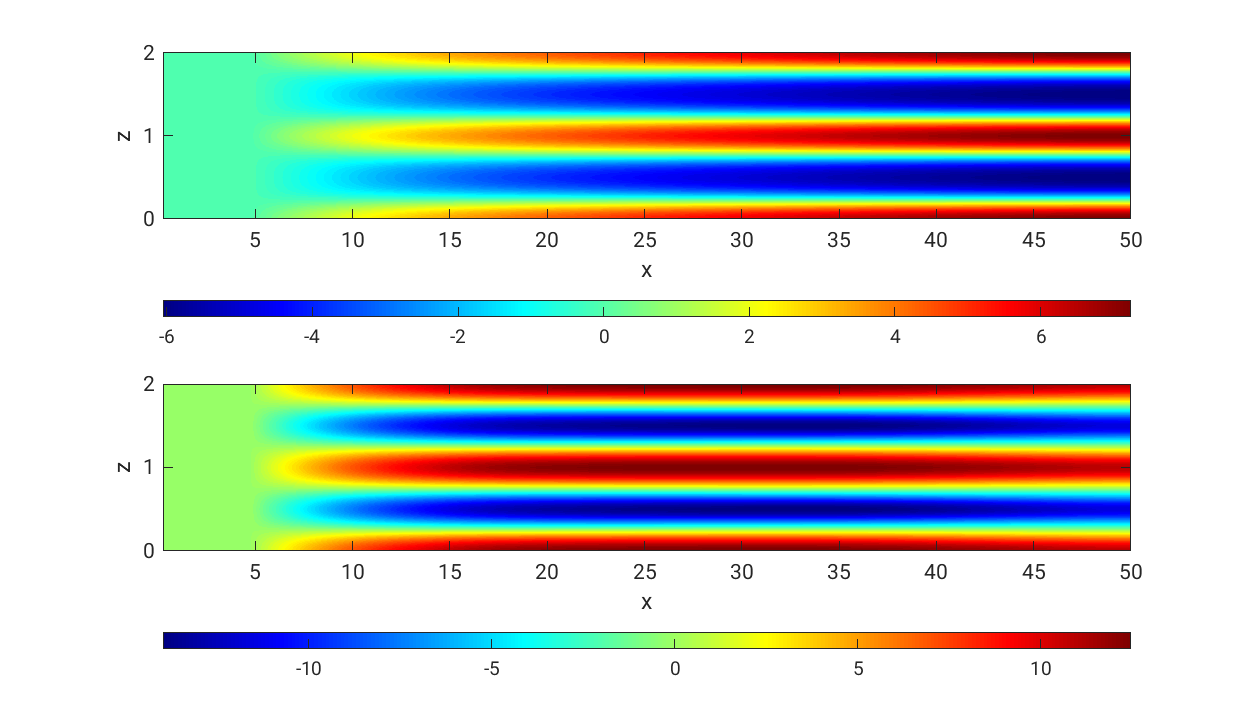}
   \includegraphics[width=8cm]{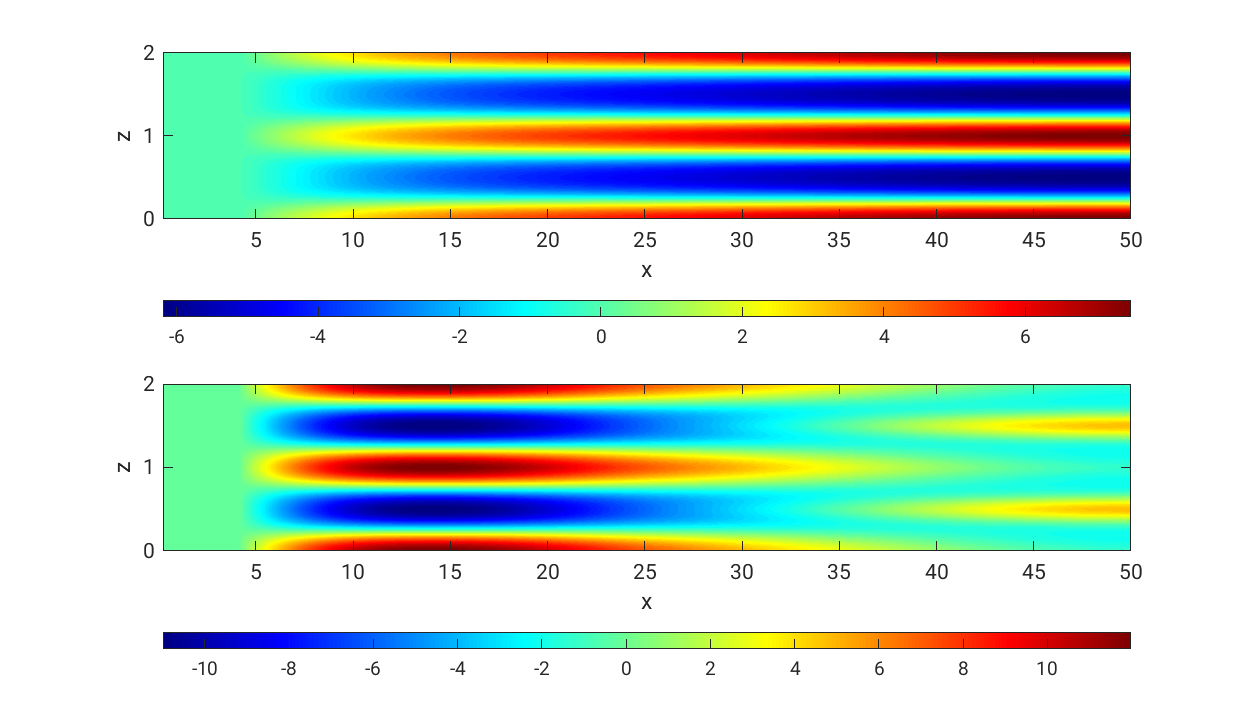}\\
   \hspace{2mm} (c) \hspace{8cm} (d)
 \end{center}
  \caption{Transpiration velocity contours in the ZX-plane for the $1^{st}$ control iteration (top) and $10^{th}$ control iteration (bottom) for $M_\infty = 3 \hspace{1mm}(a)$, $M_\infty = 4 \hspace{1mm}(b)$, $M_\infty = 5 \hspace{1mm}(c)$, and $M_\infty = 6 \hspace{1mm}(d)$}
  \label{f9}
\end{figure}

Contour plots in figure \ref{f8} show the streamwise velocity distribution in the ZX-plane for the uncontrolled (top contour) flow and the $10^{th}$ control iteration (bottom contour). We can see the alternating high- and low-speed streaks for the uncontrolled flow case. However, the contours of the controlled flow show a more uniform spanwise distribution of the streamwise velocity, indicating the disappearance of the streamwise oriented vortical structures. Figure \ref{f9} illustrates streamwise contour plots of the transpiration velocity, $v_w$. The contours show how the distribution and intensity of the control velocity vary with control iterations. 

\section{Conclusion}
In this {paper} we {conducted} an optimal control {study by formulating a novel} algorithm {that is} capable of suppressing the growth of the centrifugal instabilities {(i.e. Gortler vortices}) developing in a compressible boundary layer flow over a curved surface.
{The method involved} using the numerical solution to the nonlinear compressible boundary region equations (NCBRE), {that} a{re a} parabolized form of the Navier-Stokes equations, and Lagrange multipliers {within an} optimal control {varational problem}. 
We implemented the control mechanism of the boundary layer flow using transpiration velocity (blowing and suction) applied at the wall in the wall-normal direction {as control variables}. 
The wall shear stress {was used} as the {so-called} {'}cost-functional{'} {for} the optimal control {algorithm}. 

We applied the control {strategy} at different streamwise locations to determine the {subsequent} optimal streamwise location {that} yield{s} the highest reduction in the wall shear stress and vortex kinetic energy. As shown in figures \ref{f3} and \ref{f4}, our results demonstrate that the closer {one} move{s} the control point ($X_c$) toward the offset point {in} region III ($X_0$), the higher the decrease in the wall shear stress, and {commensurately the} vortex kinetic energy because of the increase in the control surface area. 

Figures \ref{f5} and \ref{f6} showed that the optimal control approach induces a significant reduction in wall shear stress and vortex kinetic energy as a result of the blowing or suction associated with the transpiration velocity control variable. We report $12.09\%$, $11.61\%$, $10.56\%$, and $11.02\%$ reduction in the wall shear stress and $72.31\%$, $72.38\%$, $73.27\%$, and $83.27\%$ drop in the vortex kinetic energy for respectively for the $M_\infty = 3$, $M_\infty = 4$, $M_\infty = 5$, and $M_\infty = 6$ cases when applying the control method at $X_{c_1}$ and comparing the $10^{th}$ control iteration to the uncontrolled flow case. We {have} noticed that the method requires more control iterations to achieve maximum reduction as the Mach number increases. For instance, the two high-supersonic cases ($M_\infty=3$ and  $M_\infty=4$) required six control iterations, whereas the two hypersonic ones ($M_\infty=5$ and  $M_\infty=5$) required ten. The reduction in the wall shear stress directly translates to a decrease in the boundary layer skin friction. While the reduction in the kinetic energy of the flow delays the occurrence of the secondary instabilities leading to transition, consequently elongating the laminar region of the boundary layer. Ultimately, this results in a more stable and attached boundary layer.

The crossflow contours of the streamwise primary instability in figure \ref{f7} visually captured the effect of the control approach on the centrifugal instabilities of the boundary layer. The control iterations gradually flatten out the flow instabilities (especially near the wall). 
This finding goes hand in hand with the drop in the kinetic energy and wall shear stress levels in the boundary layer. Consistent with these results, contour plots in figure \ref{f8} demonstrated how the optimal control approach results in a {more uniform} streamwise velocity distribution parallel to the wall as opposed to the high and low speed streaks of the uncontrolled flow case. 

{To conclude,} the results presented in this {paper} demonstrate that the optimal control algorithm {based on the ACBRE (adjoint compressible boundary region equations)} suppresses the growth rate of the streamwise vortex system of high supersonic and hypersonic boundary layer flows.

\section{Acknowledgments}

We would like to thank the Institute of Science at Tohoku University for the partial support. MZAK would like to thank Strathclyde University for the financial support from the Chancellor’s Fellowship.


\end{document}